\documentclass[journal,onecolumn]{IEEEtran}

\usepackage[T1]{fontenc} 
\usepackage[utf8]{inputenc}
\usepackage{pgfplots}
\usepackage{tikz}
\usepackage{pgfplots}
\pgfplotsset{compat=newest}  
\usepackage{pgfplotstable}
\usepgfplotslibrary{groupplots}
\usepackage{afterpage}
\usepackage{siunitx}
\usepackage{caption}
\ifCLASSINFOpdf
\else
\fi
\usepackage{url}

\usepackage{amsfonts}
\usepackage{amsmath}
\usepackage{subfig}
\usepackage[disable]{todonotes} 
\usepackage{bbold}
\usepackage[normalem]{ulem}
\usepackage{hyperref}

\definecolor{babyblueeyes}{rgb}{0.63, 0.79, 0.95}
\definecolor{bittersweet}{rgb}{1.0, 0.44, 0.37}
\definecolor{caribbeangreen}{rgb}{0.0, 0.8, 0.6}
\definecolor{celadon}{rgb}{0.67, 0.88, 0.69}
\definecolor{babypink}{rgb}{0.96, 0.76, 0.76}

\newcommand{\jntodo}[1]{\todo[inline, color=caribbeangreen]{@JN #1}}

\newcommand{\bctodo}[1]{\todo[inline, color=babyblueeyes]{@BĆ #1}}

\usetikzlibrary{arrows,automata}
\usetikzlibrary{positioning}

\tikzset{
	state/.style={
		rectangle,
		rounded corners,
		draw=black, very thick,
		minimum height=2em,
		inner sep=2pt,
		text centered,
	},
}

\newtheorem{re}{Remark}

\newtheorem{pr}{Proposition}
\newtheorem{df}{Definition}

\DeclareUnicodeCharacter{2212}{−}
\usepgfplotslibrary{dateplot}
\usetikzlibrary{patterns,shapes}

\hypersetup{
    colorlinks,
    linkcolor={red!80!black},
    citecolor={blue!80!black},
    urlcolor={blue!80!black}
}

\begin{document}


\title{Generalised Score Distribution: Underdispersed Continuation of the Beta-Binomial Distribution}

\author{Bogdan~Ćmiel\textsuperscript{\textsection}, Jakub~Nawała\textsuperscript{\textsection}, Lucjan~Janowski\textsuperscript{\textsection}, Krzysztof~Rusek\textsuperscript{\textsection}
    \thanks{This work is licensed under the \href{http://creativecommons.org/licenses/by/4.0/}{Creative Commons Attribution 4.0 International License}.}%
	\thanks{B.~Ćmiel is with the Department of Applied Mathematics, AGH University of Science and Technology, Poland.}
	\thanks{L.~Janowski, J.~Nawała, and K.~Rusek are with the Institute of Telecommunications of AGH University of Science and Technology, Poland. e-mail: ljanowsk@agh.edu.pl}}

\maketitle
\begingroup\renewcommand\thefootnote{\textsection}
\footnotetext{These authors contributed equally.}
\endgroup

\begin{abstract}
A class of discrete probability distributions contains distributions with limited support. A typical example is some variant of a Likert scale, with response mapped to either the $\{1, 2, \ldots, 5\}$ or $\{-3, -2, \ldots, 2, 3\}$ set.
An interesting subclass of discrete distributions with finite support are distributions limited to two parameters and having no more than one change in probability monotonicity. 
The main contribution of this paper is to propose a family of distributions fitting the above description, which we call the Generalised Score Distribution (GSD) class. The proposed GSD class covers the whole set of possible mean and variances, for any fixed and finite support. Furthermore, the GSD class can be treated as an underdispersed continuation of a reparametrized beta-binomial distribution. The GSD class parameters are intuitive and can be easily estimated by the method of moments. We also offer a Maximum Likelihood Estimation (MLE) algorithm for the GSD class and evidence that the class properly describes response distributions coming from 24 Multimedia Quality Assessment experiments.
At last, we show that the GSD class can be represented as a sum of dichotomous zero-one random variables, which points to an interesting interpretation of the class.
\end{abstract}


\section{Introduction}
\jntodo{Dopisz w abstrakcie co nas motywuje}
\jntodo{Dopasuj językowo abstrakt do wstępu}
A Likert scale is used in numerous research fields, like psychology, medicine, or quality of experience \cite{Liddell2018, AGH_NTIA_14-505}, to name a few.
Responses given to questions using a Likert scale are ordinal, but in practice they are often analysed as data coming from an interval scale. There is even a fairly common approach of ignoring the fact that data are discrete and treating them as if they come from a continuous, usually normal, distribution \cite{ITURec500}. Such analyses can result in wrong conclusions \cite{Liddell2018}.

There are many reasons why an ordinal scale is converted to an interval one for the purpose of statistical analysis. One of them is the fact that in many cases, like assessing a student's performance or rating a movie, the mean assessment (or the mean response) is used to operationalise a latent subjectively judged trait one wants to measure. For example, it is a common practice to use the average of student's grades as an assessment of their underlying capabilities. This average is treated as a measure of student's ``quality.'' Although this approach is not mathematically correct in the strict sense (assuming the data---i.e., student's grades---truly are ordinal), important decisions are made based on it (e.g., whether to grant a scholarship to a ``high quality'' student). Wanting to stay in line with how people conventionally use ordinal assessment scores (or, more generally, ordinal responses), in this work we try to model the strategy of treating quality scores as if they were expressed on an interval scale. This strategy seems justified to us especially if a trait one wants to assess has some objective characteristics that are dominant in the opinion formulation. This is the case in the field of Video Quality Assessment (VQA), where the technical reproduction quality of a visual stimulus is assessed. Although there is a subjective component to such assessments, objectively measurable factors (like resolution or the number of frames per second) are dominant in the opinion formulation. Importantly, the subjective component of the assessment can be treated as noise, distorting the measurement of the underlying stimulus quality. Our modelling approach takes that view into account.

Another reason why ordinal responses are converted to an interval scale is that proper latent variable models, like ordered logit or ordered probit (see \cite{Nelder}), are too complicated for studies with relatively few responses per hidden variable. In hidden variable models, we have the distribution parameters and the discretization/mapping parameters. All these parameters are difficult to interpret and in the case of small sample sizes, it is impossible to estimate them properly. For instance, if we have a sample with responses regarding only one or two objects (e.g., video clips), then the hidden variable models are overparameterised. They necessarily need big sample sizes with multiple raters evaluating multiple objects.

The main idea of this paper is to propose a family of discrete probability distributions that is able to model responses gathered in Multimedia Quality Assessment (MQA) subjective experiments.\footnote{Video Quality Assessment (VQA) mentioned previously is a sub-field of MQA.} There, the distribution of responses is often (but not exclusively) uderdispersed (i.e., having variance lower than that of the binomial distribution) and practitioners analysing these data need a model intuitively conveying response distribution characteristics.
We put forward a solution avoiding the difficulties inherent to treating ordinal responses as though they are expressed on an interval scale and, at the same time, offer a model with only two parameters (to sidestep overparameterisation challenges). We find our solution attractive to practitioners, who want to use relatively simple but mathematically correct tools.
We call our proposed family of distributions a \textit{Generalised Score Distribution} class (GSD). It is a two-parameter family of discrete distributions on the set $\{1,...,M\}$, $M \in \mathbb{N}\setminus\{1, 2\}$. The practical interpretation we attach to the elements of this set is that they correspond to subjective ratings of quality (e.g., of an image or a video clip). Because of its convenient parameterisation, the class does a similar job to what the normal distribution class does for the continuous case. The first parameter $\psi \in[1,M]$ is (as in the case of the normal distribution class) the expected value. In MQA, it is a common practice to assume subjective responses are expressed on an interval scale. The expected value of responses gathered in the course of a subjective experiment is then treated as an estimate of stimulus latent quality (sometimes referred to as \textit{true quality}). The $\psi$ parameter of the GSD class reflects this approach.
We would like the second parameter of the GSD class to play the same role as the second parameter of the normal distribution class. Unfortunately, for discrete distributions defined on the set $\{1,...,M\}$, the range of all possible variances is changing with $\psi$. Therefore, the second parameter $\rho\in[0,1]$ (also referred to as \textit{dispersion parameter} or \textit{confidence parameter}) of the GSD class is a linear function of variance, with the value range equal to the interval $[0,1]$ for every $\psi$. To the best of our knowledge, the convenient parameterisation of the GSD class is unique, compared to other discrete distributions. It allows us to look at the parameters describing the expected value and variance independently, as one can do in the case of the normal distribution class. In other words, shifting the expected value parameter ($\psi$), does not change the dispersion parameter ($\rho$). Likewise, changing the dispersion parameter ($\rho$), does not influence the expected value parameter ($\psi$).
The GSD class covers all possible first- and second-order moments for discrete distributions defined on $\{1,...,M\}$. It also offers explicit formulae for probabilities without the use of special functions and thus explicit formulae for the derivatives of its log-likelihood function. Therefore, we believe that the GSD can be used outside of the field of MQA. 

The problem of generalising the binomial distribution to overdispersed and underdispersed data is known and widely considered in literature. The natural generalisation of the binomial distribution for overdispersed data is the beta-binomial distribution. It provides simple formulae for probabilities and derivatives of its log-likelihood function. Furthermore, there are available ready-to-use algorithms for the estimation of its parameters (see \cite{Biometrics}). One can also find applications of the beta-binomial distribution when dealing with overdispersed data (see e.g. \cite{HospitalBBD}).
In \cite{ExtendedBBD} the method for extending the beta-binomial distribution for underdispersed data is proposed. Unfortunately, this method does not provide a distribution that covers all possible variances. For any fixed mean value, it stops at some variance and cannot go lower (see Fig. \ref{fig:varRel}). Thus, the data that are strongly underdispersed cannot be modelled in such a way. In this paper, we propose a different way of extending the beta-binomial distribution. Our solution allows to obtain all possible variances for a discrete distribution defined on $\{1,...,M\}$. Our method also provides a reparameterisation to obtain easy to interpret parameters $\psi$ (mean value) and $\rho$ (confidence level linearly dependent on variance). The GSD class can be also represented as a sum of dichotomous zero-one random variables (see Proposition \ref{GSDber}), which gives an interesting interpretation of that class.

This paper provides estimation and test of goodness-of-fit algorithms for the GSD class. We analyse the algorithms performance through an extensive simulation study. We also use six Multimedia Quality Assessment (MQA) databases (amounting to more than $100\,000$ individual responses) to provide evidence that the proposed class is a useful analytical tool that can be used in practice.
It is worth mentioning that we already proposed in the past a tool based on the GSD class. The tool extends possible ways to validate data consistency of responses obtained during a MQA subjective experiment \cite{Nawala2020ACM}. Importantly, our work was noticed by practitioners in the MQA field and referred to in \cite{Chinen2021}, \cite{Chinen2021Access}, and \cite{Hossfeld2021}.

The GSD class we propose can be used for modelling survey responses expressed on a Likert scale as well. For example, in \cite{NUMBER_OF_RESPONSE} the authors consider the impact of the number of response categories on the reliability of the measurements. In the theorised response generation model (provided in equation (1) of \cite{NUMBER_OF_RESPONSE}), one can use the GSD class to model the random error. This approach would then allow to estimate the latent unobserved true response. The same method can also be used, for example, in \cite{PSYCHOMETRIC_TESTING}, where the problem of measuring patient experience in hospitals is considered.

We claim that the GSD class properly describes responses from MQA subjective experiments. We also state that the GSD class estimates well response distributions even for samples of small size (i.e., sample sizes conventionally used in MQA experiments). At last, we argue that the GSD class describes response distributions using easy to interpret and easy to estimate parameters.
To substantiate our claims, in this work we present the following contributions:
\begin{itemize}
    \item We offer the GSD family of distributions (also referred to as the GSD class) that:
    \begin{enumerate}
        \item covers all possible first- and second-order moments for a distribution defined on a discrete finite support and
        \item uses parameterisation attractive to practitioners in the MQA field, that is similar to normal distribution’s parameterisation.
    \end{enumerate} 
    \item We evidence that the GSD family of distributions can be represented as a sum of dichotomous zero-one random variables.
    \item We show a Maximum Likelihood Estimation (MLE) algorithm for the GSD class.
    \item We indicate through goodness-of-fit testing that the GSD class well describes responses from MQA subjective experiments (in contrast to other commonly used modelling approaches).
    \item We reveal that based on samples of small size, the GSD class better forecasts a response distribution for a sample of larger size, in comparison to the empirical distribution.
\end{itemize}

The paper is structured as follows. In Section~\ref{sec:model}, we describe the proposed GSD family of distributions and compare it with the Ordered Probit model.
Section~\ref{sec:estimation} introduces the maximum likelihood estimation for the GSD class. It also considers the numerical accuracy of the estimation method we use. (Numerical accuracy was also examined in multidimensional case in Appendix \ref{app:multi}.) Section~\ref{sec:analysis} presents the analyses performed on real data sets of responses from six MQA studies. The last section concludes the paper. All proves and additional formulae can be found in Appendices.


\section{Subjective Response as a Random Variable}
\label{sec:model}

Assuming that we use an $M$-point discrete scale, a random variable $U$ (describing a subjective response) has a distribution given by:
\begin{equation}
P(U = s) = p_{s}, \textrm{ where } \sum_{s=1}^{M}p_s = 1
\end{equation}
Such a description of a response distribution is general but has $M-1$ different parameters. There are $M-1$ of them, since there are $M$ probabilities this distribution describes. 
In general, we can describe a subjective response as a function: 
\begin{equation}
U = \psi + \epsilon,
\end{equation}
where $\psi$ is the expected value (referred to as true quality\footnotetext{Our notation convention generally follows the guidelines of VQEG (Video Quality Expert Group), described in \cite{Janowski2019NotationFS}.} in the context of MQA research) and $\epsilon$ is an error term with the mean value equal to zero. An algorithm predicting stimulus quality (or any other subjectively judged trait) should aim at estimating $\psi$. Still, the error distribution is important and should be modelled. The error term represents the precision of $\psi$ estimation. It is desirable that the error term (represented by $\epsilon$) should not be too complicated. Therefore, we would like to use a model in which the error is described by a single parameter. (Please note that in Appendix \ref{app:multi} we also consider a multidimensional version of subjective responses with $m$ quality parameters and $n$ error parameters.)


\subsection{Ordered Probit With Fixed Thresholds}

The models proposed by \cite{Janowski2015} and \cite{Li2017} describe subjective responses as following a continuous normal distribution with certain mean $\mu$, which is assumed to represent the latent stimulus quality (also referred to as true quality), and standard deviation $\sigma$, describing the error. Therefore, subjective response $O \sim \mathcal{N}(\mu, \sigma^2)$. Since in MQA experiments subjective responses are often expressed on a discrete scale, we cannot directly observe $O$. To convert the continuous form of a response to a discrete one, discretisation and censoring (clipping) are necessary. This process converts a continuous random variable $O$ to a discrete variable $U$. We can calculate each response category probability (i.e., $U$ distribution), as a function of $\mu$ and $\sigma$ using the following equations:
\begin{equation}
P(U = s) = \int_{s-0.5}^{s+0.5}\frac{1}{2\pi \sigma}e^{-\frac{(x-\mu)^2}{2\sigma^2}} 
\end{equation}
for $s = \{2,3,M-1\}$ and 
\begin{equation}
P(U = 1) = \int_{-\infty}^{1.5}\frac{1}{2\pi \sigma}e^{-\frac{(x-\mu)^2}{2\sigma^2}}, \ \ \ P(U = M) = \int_{M-0.5}^{\infty}\frac{1}{2\pi \sigma}e^{-\frac{(x-\mu)^2}{2\sigma^2}}.
\end{equation}

\begin{figure}
	\centering
	\input{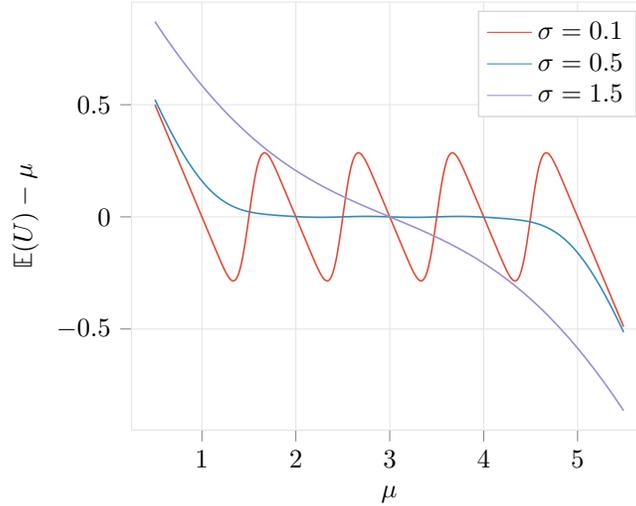}
	\caption{The difference between $\mathbb{E}(U)$ and $\mu$ (the ordered probit parameter) for $M=5$ and different $\sigma$.}
	\label{fig:smallvar}
\end{figure}

Note that the definition of true quality $\mu$ is model dependent here. This is a flaw of this approach. It is also worth mentioning that $\mu=\mathbb{E}(O)$ can be completely different from $\psi=\mathbb{E}(U)$. The latter is a natural and model-independent parameter defining the true quality (for possible differences between $\mathbb{E}(O)$ and $\mathbb{E}(U)$ see Fig.~\ref{fig:smallvar}).

It is obvious that $\mathbb{E}(O)\in{(-\infty,\infty)}$ and $\mathbb{E}(U)\in[1,M]$ have to be different. However, even for $\mathbb{E}(O)\in[1,M]$ the differences are considerable. The other problem, especially from the numerical estimation point of view, is the unbounded parameter set $(\mu,\sigma)\in(-\infty,\infty)\times(0,\infty)$.
In Fig.~\ref{fig:ghost_ordered_probit} one can see how parameters $(\mu,\sigma)$ map to $(\mathbb{E}(U),\mathbb{V}(U))$ after discretisation, for the case $M=5$.
This figure and the lack of an inverse formula for calculating $(\mu,\sigma)$ having $(\mathbb{E}(U),\mathbb{V}(U))$, make it clear that this approach (referred to in the literature as \textit{ordered probit}~\cite{Becker1992}) may prove problematic if the moments-based estimation would be used. Likewise, it may be challenging to find a starting point for numerical estimation methods.





\subsection{GSD}
\label{ssec:gsd}
An example of a discrete distribution that is described by two parameters is the beta-binomial distribution \cite{betaBinomial}. The smallest variance the beta-binomial distribution can express is binomial distribution's variance. It is a strong limitation. For example, in the case of MQA subjective experiments for video, in \cite{Hossfeld2018} it is suggested that the binomial distribution has the highest possible variance for a correctly conducted subjective experiment. Differently put, most MQA subjective experiments yield data with response distributions having the variance lower than that of the binomial distribution. Therefore, we need a different distribution, covering the whole spectrum of possible variances. This is especially true for underdispersed probability distributions (i.e., distributions with the variance lower than that of the binomial distribution; see Fig. \ref{fig:varRel}). 


\subsubsection{GSD Construction and Definition}
\label{sssec:gsd_construction_and_definition}
Let us start with the equation describing subjective responses 
\begin{equation}
U = \psi + \epsilon,
\end{equation}
where $\epsilon$ is an error with mean value equal to $0$. Since $U$ belongs to the set $\{1,2,...,M\}$, then the distribution of $\epsilon$ has to be supported on the set $1-\psi$, $2-\psi$, ..., $M-\psi$. 
Let us consider the shifted binomial distribution for $\epsilon$:
$$P(\epsilon=k-\psi)=\binom{M-1}{k-1}\left(\frac{\psi-1}{M-1}\right)^{k-1}\left(\frac{M-\psi}{M-1}\right)^{M-k},$$
where $k\in\{1,...,M\}$ represents the response categories, from which subjective experiment participants (also referred to as \textit{subjects} or \textit{raters}) can choose from. 

\begin{figure}
	\centering
	\includegraphics[width=0.7\textwidth]{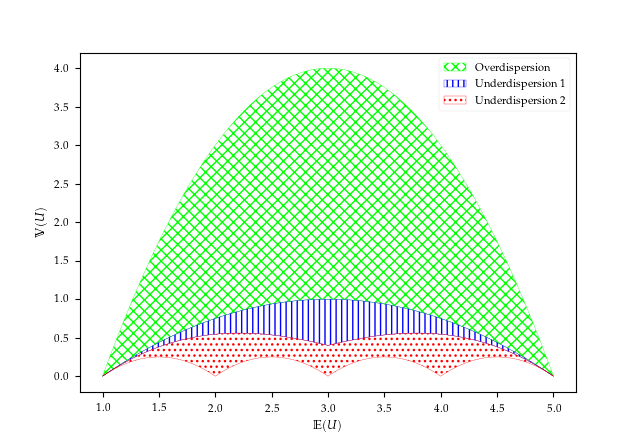}
	\caption{Area of all possible $(\mathbb{E}(U),\mathbb{V}(U))$ for discrete distributions on $\{1,...,M\}$ for $M=5$. The area is divided into different parts covered by two-parameter discrete models: the beta-binomial distribution (green crosses), extended beta-binomial distribution (previous + blue vertical hatching), and our solution (covers all possible values including the area marked with red dots).}
	\label{fig:varRel}
\end{figure}

Since the support of this distribution and the mean value are fixed, we obtain a fixed shifted binomial distribution without any freedom. However, we would like to have a class of distributions for $\epsilon$ with all possible variances $\mathbb{V}(\epsilon)=\mathbb{V}(U)$. Let us think about how the set of all possible variances, for all distributions supported on the set $1-\psi$, $2-\psi$, ..., $M-\psi$, looks like.
Remember that the mean values for such distributions are fixed at $0$ (since they describe the error term). This is why the set of all possible variances depends on $\psi$. If we denote by $V_{\mathrm{min}}(\psi), V_{\mathrm{max}}(\psi)$ the minimal and maximal possible variance, respectively, then
\begin{equation}\label{Varmin}
V_{\mathrm{min}}(\psi)=(\lceil\psi\rceil-\psi)(\psi-\lfloor\psi\rfloor),
\end{equation}
\begin{equation}\label{Varmax}
V_{\mathrm{max}}(\psi)=(\psi-1)(M-\psi),
\end{equation}
and the interval $[V_{\mathrm{min}}(\psi),V_{\mathrm{max}}(\psi)]$ is the set of all possible variances. Notice that the interval $[V_{\mathrm{min}}(\psi),V_{\mathrm{max}}(\psi)]$ is the biggest for $\psi=(M+1)/2$ and if $\psi$ is not an integer, then $V_{\mathrm{min}}(\psi)>0$.
Let us return to the shifted binomial distribution. It is easy to calculate that its variance is equal to: 
\begin{equation}\label{eq:VarBin}
V_{\mathrm{Bin}}(\psi):=\frac{V_{\mathrm{max}}(\psi)}{M-1}.
\end{equation}
The question is how to obtain from this shifted binomial distribution a class of distributions that covers the whole interval of variances $[V_{\mathrm{min}}(\psi),V_{\mathrm{max}}(\psi)]$ for any $\psi$ (see Fig. \ref{fig:varRel}). 

\begin{figure}
	\begin{center}
		\input{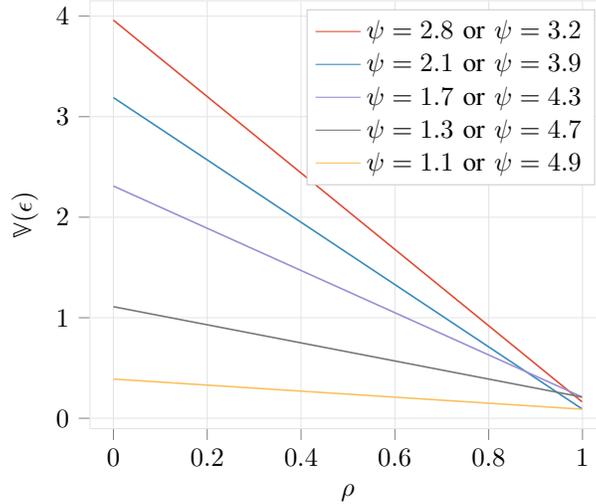}
		\caption{Variance of the error term $\epsilon$ for $M=5$.} \label{fig:varMOS}
	\end{center}
\end{figure}

We would like to obtain this class by:
\begin{itemize}
	\item adding only a single normalised parameter $\rho\in[0,1]$, 
	\item making variance of the error to be linearly dependent on $\rho$ and
	\item requiring that variance is a decreasing function of $\rho$ (this way we could interpret $\rho$ as a confidence parameter; see Fig. \ref{fig:varMOS}). 
\end{itemize}

Let us denote by $H_{\rho}$ the distribution of the error fulfilling the above conditions. Since the variance of the error is linearly dependent on $\rho$ and decreasing, then it has to be equal to
\begin{equation}\label{eq:Var}
V_{H_{\rho}}(\epsilon)=\rho V_{\mathrm{min}}(\psi)+(1-\rho)V_{\mathrm{max}}(\psi).
\end{equation}  
Using formulae (\ref{eq:VarBin}) and (\ref{eq:Var}) we can calculate 
\begin{equation}\label{Cpsi}
V_{H_{\rho}}(\epsilon)=V_{\mathrm{Bin}}(\psi) \  \Leftrightarrow  \ \rho=C(\psi):=\frac{M-2}{M-1}\ \frac{V_{\mathrm{max}}(\psi)}{V_{\mathrm{max}}(\psi)-V_{\mathrm{min}}(\psi)},
\end{equation}
which gives us the value of $\rho$ corresponding to a shifted binomial distribution. We have 
$$
V_{H_{\rho}}(\epsilon)\in [V_{\mathrm{min}}(\psi),V_{\mathrm{Bin}}(\psi)] \Leftrightarrow \rho \in [C(\psi),1],
$$
which corresponds to the red coloured dots and blue vertical hatching in Fig. \ref{fig:varRel}, and
$$
V_{H_{\rho}}(\epsilon)\in [V_{\mathrm{Bin}}(\psi),V_{\mathrm{max}}(\psi)] \Leftrightarrow \rho \in [0,C(\psi)],
$$
which corresponds to the green coloured area in Fig. \ref{fig:varRel}. 

For variances bigger than $V_{\mathrm{Bin}}(\psi)$ (the green coloured area in Fig. \ref{fig:varRel}) we use the reparameterised beta binomial distribution. Since the mean value is fixed, we only have one free parameter $\rho\in[0,C(\psi)]$. The effect of such reparameterization gives us the distribution denoted by $G_{\rho}$:
\begin{equation}
	P_{G_{\rho}} (\epsilon=k-\psi)	= \binom{M-1}{k-1}
	\frac{\prod\limits_{i=0}^{k-2}\left(\frac{(\psi-1)\rho}{(M-1)}+i(C(\psi)-\rho)\right)\prod\limits_{j=0}^{M-k-1}\left(\frac{(M-\psi)\rho}{(M-1)}+j(C(\psi)-\rho)\right)}{\prod\limits_{i=0}^{M-2}\left(\rho+i(C(\psi)-\rho)\right)}, 
	\label{eq:Prho}
\end{equation}
where $\rho\in[0,C(\psi)]$ and $k\in\{1,...,M\}$. The above formula can be rewritten as
$$
	P_{G_{\rho}} (\epsilon=k-\psi)	=\begin{cases} 
	\frac{M-\psi}{M-1}\prod\limits_{i=1}^{M-2} \frac{\frac{(M-\psi)\rho}{(M-1)}+i(C(\psi)-\rho)}{\rho+i(C(\psi)-\rho)} \ \ \mathrm{for} \ \ k=1 \\ 
	\binom{M-1}{k-1} \frac{(\psi-1)(M-\psi)\rho}{(M-1)^2}
	\frac{\prod\limits_{i=1}^{k-2}\left(\frac{(\psi-1)\rho}{(M-1)}+i(C(\psi)-\rho)\right)\prod\limits_{j=1}^{M-k-1}\left(\frac{(M-\psi)\rho}{(M-1)}+j(C(\psi)-\rho)\right)}{\prod\limits_{i=1}^{M-2}\left(\rho+i(C(\psi)-\rho)\right)} \ \ \mathrm{for} \ \ k=2,...,M-1\\
	\frac{\psi-1}{M-1}\prod\limits_{i=1}^{M-2} \frac{\frac{(\psi-1)\rho}{(M-1)}+i(C(\psi)-\rho)}{\rho+i(C(\psi)-\rho)} \ \ \mathrm{for} \ \ k=M
	\end{cases}
$$
Therefore, for $\rho=0$ we obtain
$$
	P_{G_{0}} (\epsilon=k-\psi)	=\begin{cases} 
	\frac{M-\psi}{M-1} \ \ \mathrm{for} \ \ k=1 \\ 
	\ \ \ 0 \ \ \ \ \mathrm{for} \ \ k=2,...,M-1\\
    \frac{\psi-1}{M-1} \ \ \mathrm{for} \ \ k=M
	\end{cases}
$$
\begin{pr}\label{GSDbetabinomial}
If $\epsilon$ has $G_{\rho}$ distribution for fixed $\psi\in[1,M]$ and $\rho\in[0,C(\psi)]$, then 
$$\mathbb{E}(U)=\psi, \ \ \mathbb{V}(U)=\rho V_{\mathrm{min}}(\psi)+(1-\rho)V_{\mathrm{max}}(\psi), \ \ \mathbb{V}(U)\in[V_{\mathrm{Bin}}(\psi),V_{\mathrm{max}}(\psi)],$$
where $U=\psi+\epsilon$ is supported on $\{1,...,M\}$.  
\end{pr}
The proof of Proposition \ref{GSDbetabinomial} can be found in Appendix \ref{proofs}.

\begin{re}
	Notice that for $\rho \rightarrow 0$ the $G_{\rho}$ distribution approaches a two-point distribution supported on $\{1-\psi,M-\psi\}$, with the biggest possible variance equal to $V_{\mathrm{max}}(\psi)$. For $\rho \rightarrow C(\psi)$ the $G_{\rho}$ distribution approaches the shifted binomial distribution, with variance equal to $V_{\mathrm{Bin}}(\psi)$.
\end{re}

For variances smaller than $V_{\mathrm{Bin}}(\psi)$ (cf. the blue vertical hatching and red coloured dots in Fig. \ref{fig:varRel}) we use a mixture technique. Specifically, we take a mixture of the shifted binomial distribution and the distribution with the smallest possible variance (i.e., a two-point or one-point distribution, depending on $\psi$). Of course, the mixture parameter has to be reparameterised to fit the $[C(\psi),1]$ interval.
The effect of such reparameterisation gives us the distribution denoted by $F_{\rho}$:
\begin{equation} 
		P_{F_\rho}  (\epsilon=k-\psi)= 
		 \frac{\rho-C(\psi)}{1-C(\psi)}[1-|k-\psi|]_{+} + \frac{1-\rho}{1-C(\psi)} 
		 \binom{M-1}{k-1}\left(\frac{\psi-1}{M-1}\right)^{k-1}\left(\frac{M-\psi}{M-1}\right)^{M-k},
 \label{eq:Frho}
\end{equation}
where $\rho\in[C(\psi),1]$, $[x]_{+} = \max(x,0)$ and $k\in\{1,...,M\}$.
\begin{pr}\label{GSDmix}
If $\epsilon$ has $F_{\rho}$ distribution for fixed $\psi\in[1,M]$ and $\rho\in[C(\psi),1]$, then 
$$\mathbb{E}(U)=\psi, \ \ \mathbb{V}(U)=\rho V_{\mathrm{min}}(\psi)+(1-\rho)V_{\mathrm{max}}(\psi), \ \ \mathbb{V}(U)\in[V_{\mathrm{min}}(\psi),V_{\mathrm{Bin}}(\psi)],$$
where $U=\psi+\epsilon$ is supported on $\{1,...,M\}$. 
\end{pr}
The proof of Proposition \ref{GSDmix} can be found in Appendix \ref{proofs}.
\begin{re}
	Notice that for $\rho \rightarrow C(\psi)$ the distribution $F_{\rho}$ approaches the shifted binomial distribution with variance equal to $V_{\mathrm{Bin}}(\psi)$.
	For $\rho \rightarrow 1$ the distribution $F_{\rho}$ approaches a two-point or one-point distribution (depending on $\psi$), with the smallest possible variance equal to $V_{\mathrm{min}}(\psi)$.
\end{re}
Finally, we obtain the distribution:
$$H_{\rho}=G_{\rho}\ I(\rho<C(\psi)) + F_{\rho}\ I(\rho\geq C(\psi)).$$
\begin{df}
If $\epsilon$ has $H_{\rho}$ distribution for fixed $\psi\in[1,M]$ and $\rho\in[0,1]$, then we say that $U=\psi+\epsilon$ has the GSD($\psi,\rho$) distribution, where $\psi$ is the expected value and $\rho\in[0,1]$ is a confidence parameter, linearly dependent on the variance, i.e.,
$$
\rho=\frac{V_{\mathrm{max}}(\psi)-\mathbb{V}(U)}{V_{\mathrm{max}}(\psi)-V_{\mathrm{min}}(\psi)}
$$
\end{df}
\begin{pr}\label{GSD}
If $U$ supported on $\{1,...,M\}$ has the GSD($\psi,\rho$) distribution for $\psi\in[1,M]$ and $\rho\in[0,1]$, then  
$$\mathbb{E}(U)=\psi, \ \ \mathbb{V}(U)=\rho V_{\mathrm{min}}(\psi)+(1-\rho)V_{\mathrm{max}}(\psi), \ \ \mathbb{V}(U)\in[V_{\mathrm{min}}(\psi),V_{\mathrm{max}}(\psi)].$$
\end{pr}
The proof of Proposition \ref{GSD} is an obvious consequence of Propositions \ref{GSDbetabinomial} and \ref{GSDmix}.

\begin{figure}
	\begin{center}
\begin{tikzpicture}

\definecolor{color0}{rgb}{0.12156862745098,0.466666666666667,0.705882352941177}
\definecolor{color1}{rgb}{1,0.498039215686275,0.0549019607843137}

\begin{axis}[
tick align=outside,
tick pos=left,
x grid style={white!69.0196078431373!black},
xlabel={\(\displaystyle x\)},
xmin=0.8, xmax=4.2,
xtick style={color=black},
xtick={1,2,2.3,3,3.3,4},
xticklabels={1,2,\(\displaystyle \psi-1\),3,\(\displaystyle \psi\),4},
y grid style={white!69.0196078431373!black},
ylabel={\(\displaystyle \phi_{\psi,\rho}(x)\)},
ymin=0, ymax=1,
ytick style={color=black},
ytick={0,0.277198697068404,0.795114006514658,1},
yticklabels={
  0,
  \(\displaystyle \frac{(1 - \rho)(M - \psi)}{(1 - C(\psi))(M - 1)}\),
  \(\displaystyle 1 - \frac{(1 - \rho)(\psi - 1)}{(1 - C(\psi))(M - 1)}\),
  1
}
]
\path [draw=black, very thin, dash pattern=on 1.85pt off 0.8pt]
(axis cs:1,0)
--(axis cs:1,1);

\path [draw=black, very thin, dash pattern=on 1.85pt off 0.8pt]
(axis cs:2,0)
--(axis cs:2,1);

\path [draw=black, very thin, dash pattern=on 1.85pt off 0.8pt]
(axis cs:2.3,0)
--(axis cs:2.3,1);

\path [draw=black, very thin, dash pattern=on 1.85pt off 0.8pt]
(axis cs:3,0)
--(axis cs:3,1);

\path [draw=black, very thin, dash pattern=on 1.85pt off 0.8pt]
(axis cs:3.3,0)
--(axis cs:3.3,1);

\path [draw=black, very thin, dash pattern=on 1.85pt off 0.8pt]
(axis cs:4,0)
--(axis cs:4,1);

\path [draw=black, very thin, dash pattern=on 1.85pt off 0.8pt]
(axis cs:0.8,0.277198697068404)
--(axis cs:4,0.277198697068404);

\path [draw=black, very thin, dash pattern=on 1.85pt off 0.8pt]
(axis cs:0.8,0.795114006514658)
--(axis cs:1,0.795114006514658);

\addplot [line width=0.56pt, color0]
table {%
1 0.795114006514658
2 0.795114006514658
2.3 0.795114006514658
3 0.43257328990228
3.3 0.277198697068404
4 0.277198697068404
};
\addplot [line width=0.56pt, color1, mark=x, mark size=6, mark options={solid}, only marks]
table {%
1 0.795114006514658
2 0.795114006514658
3 0.43257328990228
4 0.277198697068404
};
\draw (axis cs:3.4,0.3) node[
  scale=0.5,
  anchor=base west,
  text=black,
  rotate=0.0
]{$\phi_{\psi,\rho}(x)$};
\end{axis}

\end{tikzpicture}
		\caption{Example of $\phi_{\psi,\rho}(x)$ for $\psi=3.3$, $\rho=0.9$, and $M=5$.} \label{fig:phi}
	\end{center}
\end{figure}
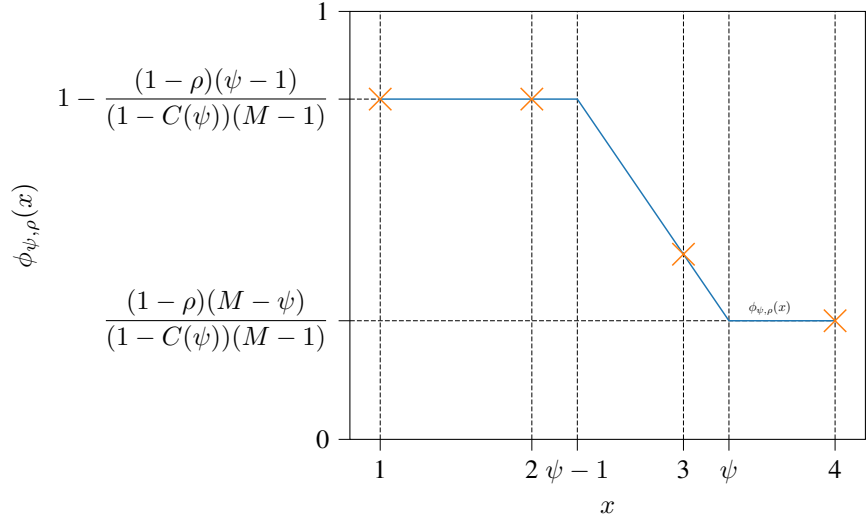

In the following proposition, we show that our GSD class can be looked at from an interesting angle. The GSD class can be represented as a distribution of the number of successes in a sequence of $M-1$ experiments.
\begin{pr}\label{GSDber}
GSD distribution can be represented as a sum of dichotomous zero-one random variables. Specifically, $U=1+\sum\limits_{i=1}^{M-1} Z_i$ has the GSD($\psi,\rho$) distribution if:
\begin{enumerate}
    \item[a)] in the case of $\rho\geq C(\psi)$, $Z_1,...,Z_{M-1}$ are zero-one independent random variables and
$$P(Z_i=1)=\phi_{\psi,\rho}(i),$$
where (see Fig. \ref{fig:phi})
$$
\phi_{\psi,\rho}(x)=\begin{cases} \frac{\rho-C(\psi)}{1-C(\psi)}+\frac{(1-\rho)(\psi-1)}{(1-C(\psi))(M-1)} \ \ \mathrm{for}\ \ x\leq \psi-1\\
\frac{\rho-C(\psi)}{1-C(\psi)}(\psi-x)+\frac{(1-\rho)(\psi-1)}{(1-C(\psi))(M-1)} \ \ \mathrm{for}\ \ x\in(\psi-1,\psi] \\
\frac{(1-\rho)(\psi-1)}{(1-C(\psi))(M-1)} \ \ \mathrm{for}\ \ x>\psi \end{cases}.
$$
    \item[b)] in the case of $\rho<C(\psi)$, $Z_1|B,...,Z_{M-1}|B$ are conditionally independent zero-one random variables, where $P(Z_i=1|B)=B$ and $B$ has the beta distribution of the following form: $\mathcal{B}\left(\frac{(\psi-1)\rho}{(M-1)(C(\psi)-\rho)},\frac{(M-\psi)\rho}{(M-1)(C(\psi)-\rho)}\right)$.
\end{enumerate}
\end{pr}
The proof of Proposition \ref{GSDber} can be found in Appendix \ref{proofs}.

\begin{re}
If we consider a class of functions $\phi_{\psi,\rho}$ satisfying the following conditions:
\begin{itemize}
    \item $\forall\ x\in[1,M-1]\ \  \phi_{\psi,\rho}(x)\in[0,1]$,
    \item  $\sum\limits_{i=1}^{M-1} \phi_{\psi,\rho}(i)=\psi-1$,
    \item  $\sum\limits_{i=1}^{M-1} [\phi_{\psi,\rho}(i)]^2=\psi-1-(1-\rho)V_{\mathrm{max}}(\psi)-\rho V_{\mathrm{min}}(\psi)$,
\end{itemize}
then we obtain a completely general response distribution suitable for representing underdispersed data (cf. Fig.~\ref{fig:varRel}) with mean value $\psi$ and confidence parameter $\rho$.
\end{re}

\begin{figure*}
	\begin{center}
		\subfloat{
\begin{tikzpicture}

\definecolor{color0}{rgb}{0.886274509803922,0.290196078431373,0.2}
\definecolor{color1}{rgb}{0.203921568627451,0.541176470588235,0.741176470588235}
\definecolor{color2}{rgb}{0.596078431372549,0.556862745098039,0.835294117647059}
\definecolor{color3}{rgb}{0.984313725490196,0.756862745098039,0.368627450980392}
\definecolor{color4}{rgb}{0.556862745098039,0.729411764705882,0.258823529411765}
\definecolor{color5}{rgb}{0.75,0,0.75}

\begin{axis}[
axis background/.style={fill=white},
axis line style={white!89.80392156862746!black},
legend cell align={left},
legend entries={\hspace{-.2cm}\textbf{$\rho$},{0.95},{0.88},{0.81},{0.72},{0.61},{0.38}},
legend style={draw=white!80.0!black, fill=white},
tick align=outside,
tick pos=left,
x grid style={white!89.80392156862746!black},
xlabel={Score $s$},
xmajorgrids,
xmin=0.8, xmax=5.2,
y grid style={white!89.80392156862746!black},
ylabel={$P(U = s)$},
ymajorgrids,
ymin=-0.05, ymax=1.05
]
\addlegendimage{empty legend}
\addplot [line width=0.08000000000000002pt, color0, dashed, mark=*, mark size=4, mark options={solid}]
table [row sep=\\]{%
1	0.72139609375 \\
2	0.258290625 \\
3	0.0192515625 \\
4	0.001040625 \\
5	2.109375e-05 \\
};
\addplot [line width=0.08000000000000002pt, color1, dashed, mark=triangle*, mark size=4, mark options={solid,rotate=180}]
table [row sep=\\]{%
1	0.748496941750038 \\
2	0.208222542278198 \\
3	0.038350722665177 \\
4	0.00464316083489919 \\
5	0.000286632471687573 \\
};
\addplot [line width=0.08000000000000002pt, color2, dashed, mark=diamond*, mark size=4, mark options={solid}]
table [row sep=\\]{%
1	0.77096417921922 \\
2	0.171207946584666 \\
3	0.0460910849306944 \\
4	0.0103372735077423 \\
5	0.00139951575767982 \\
};
\addplot [line width=0.08000000000000002pt, white!46.666666666666664!black, dashed, mark=x, mark size=4, mark options={solid}]
table [row sep=\\]{%
1	0.795829157800537 \\
2	0.134191333399622 \\
3	0.0484510529979119 \\
4	0.017207262603162 \\
5	0.00432119319876703 \\
};
\addplot [line width=0.08000000000000002pt, color3, dashed, mark=pentagon*, mark size=4, mark options={solid}]
table [row sep=\\]{%
1	0.821895746624118 \\
2	0.0996567312454621 \\
3	0.0451553265189036 \\
4	0.0231361667293331 \\
5	0.0101560288821829 \\
};
\addplot [line width=0.08000000000000002pt, color4, dashed, mark=square*, mark size=4, mark options={solid}]
table [row sep=\\]{%
1	0.866684689952905 \\
2	0.0497367503924647 \\
3	0.0296816091051805 \\
4	0.0246877708006279 \\
5	0.0292091797488226 \\
};
\addplot [very thick, color5, forget plot]
table [row sep=\\]{%
1.3	0 \\
1.3	1 \\
};
\node at (axis cs:1.4,0.8)[
  anchor=base west,
  text=black,
  rotate=0.0
]{ $\psi = 1.30$};
\end{axis}

\end{tikzpicture}}
		\subfloat{
\begin{tikzpicture}

\definecolor{color0}{rgb}{0.886274509803922,0.290196078431373,0.2}
\definecolor{color1}{rgb}{0.203921568627451,0.541176470588235,0.741176470588235}
\definecolor{color2}{rgb}{0.596078431372549,0.556862745098039,0.835294117647059}
\definecolor{color3}{rgb}{0.984313725490196,0.756862745098039,0.368627450980392}
\definecolor{color4}{rgb}{0.556862745098039,0.729411764705882,0.258823529411765}
\definecolor{color5}{rgb}{0.75,0,0.75}

\begin{axis}[
axis background/.style={fill=white},
axis line style={white!89.80392156862746!black},
legend cell align={left},
legend entries={\hspace{-.2cm}\textbf{$\rho$},{0.95},{0.88},{0.81},{0.72},{0.61},{0.38}},
legend style={draw=white!80.0!black, fill=white},
tick align=outside,
tick pos=left,
x grid style={white!89.80392156862746!black},
xlabel={Score $s$},
xmajorgrids,
xmin=0.8, xmax=5.2,
y grid style={white!89.80392156862746!black},
ylabel={$P(U = s)$},
ymajorgrids,
ymin=-0.05, ymax=1.05
]
\addlegendimage{empty legend}
\addplot [line width=0.08000000000000002pt, color0, dashed, mark=*, mark size=4, mark options={solid}]
table [row sep=\\]{%
1	0.0605281332818022 \\
2	0.794662643551237 \\
3	0.130343269655477 \\
4	0.0132129969081272 \\
5	0.00125295660335689 \\
};
\addplot [line width=0.08000000000000002pt, color1, dashed, mark=triangle*, mark size=4, mark options={solid,rotate=180}]
table [row sep=\\]{%
1	0.145267519876325 \\
2	0.647190344522968 \\
3	0.172823847173145 \\
4	0.0317111925795053 \\
5	0.00300709584805654 \\
};
\addplot [line width=0.08000000000000002pt, color2, dashed, mark=diamond*, mark size=4, mark options={solid}]
table [row sep=\\]{%
1	0.230006906470848 \\
2	0.4997180454947 \\
3	0.215304424690813 \\
4	0.0502093882508834 \\
5	0.00476123509275618 \\
};
\addplot [line width=0.08000000000000002pt, white!46.666666666666664!black, dashed, mark=x, mark size=4, mark options={solid}]
table [row sep=\\]{%
1	0.316802044146678 \\
2	0.37043976700906 \\
3	0.221868434092756 \\
4	0.0777356542005991 \\
5	0.0131541005509078 \\
};
\addplot [line width=0.08000000000000002pt, color3, dashed, mark=pentagon*, mark size=4, mark options={solid}]
table [row sep=\\]{%
1	0.394171811991077 \\
2	0.285141337039146 \\
3	0.184045116936097 \\
4	0.0997985070460573 \\
5	0.0368432269876218 \\
};
\addplot [line width=0.08000000000000002pt, color4, dashed, mark=square*, mark size=4, mark options={solid}]
table [row sep=\\]{%
1	0.532202936822452 \\
2	0.153334274334812 \\
3	0.107779556060854 \\
4	0.0956263175840495 \\
5	0.111056915197833 \\
};
\addplot [very thick, color5, forget plot]
table [row sep=\\]{%
2.1	0 \\
2.1	1 \\
};
\node at (axis cs:2.2,0.8)[
  anchor=base west,
  text=black,
  rotate=0.0
]{ $\psi = 2.10$};
\end{axis}

\end{tikzpicture}}
		\\
		\vspace*{-0.3cm}
		\subfloat{
\begin{tikzpicture}

\definecolor{color0}{rgb}{0.886274509803922,0.290196078431373,0.2}
\definecolor{color1}{rgb}{0.203921568627451,0.541176470588235,0.741176470588235}
\definecolor{color2}{rgb}{0.596078431372549,0.556862745098039,0.835294117647059}
\definecolor{color3}{rgb}{0.984313725490196,0.756862745098039,0.368627450980392}
\definecolor{color4}{rgb}{0.556862745098039,0.729411764705882,0.258823529411765}
\definecolor{color5}{rgb}{0.75,0,0.75}

\begin{axis}[
axis background/.style={fill=white},
axis line style={white!89.80392156862746!black},
legend cell align={left},
legend entries={\hspace{-.2cm}\textbf{$\rho$},{0.95},{0.88},{0.81},{0.72},{0.61},{0.38}},
legend style={draw=white!80.0!black, fill=white},
tick align=outside,
tick pos=left,
x grid style={white!89.80392156862746!black},
xlabel={Score $s$},
xmajorgrids,
xmin=0.8, xmax=5.2,
y grid style={white!89.80392156862746!black},
ylabel={$P(U = s)$},
ymajorgrids,
ymin=-0.05, ymax=1.05
]
\addlegendimage{empty legend}
\addplot [line width=0.08000000000000002pt, color0, dashed, mark=*, mark size=4, mark options={solid}]
table [row sep=\\]{%
1	0.0185348096301595 \\
2	0.18048492784562 \\
3	0.743586358682183 \\
4	0.0472332605781363 \\
5	0.0101606432639014 \\
};
\addplot [line width=0.08000000000000002pt, color1, dashed, mark=triangle*, mark size=4, mark options={solid,rotate=180}]
table [row sep=\\]{%
1	0.0444835431123828 \\
2	0.223163826829488 \\
3	0.594607260837239 \\
4	0.113359825387527 \\
5	0.0243855438333634 \\
};
\addplot [line width=0.08000000000000002pt, color2, dashed, mark=diamond*, mark size=4, mark options={solid}]
table [row sep=\\]{%
1	0.0704322765946062 \\
2	0.265842725813356 \\
3	0.445628162992295 \\
4	0.179486390196918 \\
5	0.0386104444028254 \\
};
\addplot [line width=0.08000000000000002pt, white!46.666666666666664!black, dashed, mark=x, mark size=4, mark options={solid}]
table [row sep=\\]{%
1	0.114405372370277 \\
2	0.276308916717531 \\
3	0.323641015625745 \\
4	0.216169729114808 \\
5	0.0694749661716385 \\
};
\addplot [line width=0.08000000000000002pt, color3, dashed, mark=pentagon*, mark size=4, mark options={solid}]
table [row sep=\\]{%
1	0.178081405530565 \\
2	0.244367463186289 \\
3	0.249159177257745 \\
4	0.206253633803385 \\
5	0.122138320222017 \\
};
\addplot [line width=0.08000000000000002pt, color4, dashed, mark=square*, mark size=4, mark options={solid}]
table [row sep=\\]{%
1	0.313469658678537 \\
2	0.158680186692613 \\
3	0.136641487850938 \\
4	0.146797829506136 \\
5	0.244410837271775 \\
};
\addplot [very thick, color5, forget plot]
table [row sep=\\]{%
2.85	0 \\
2.85	1 \\
};
\node at (axis cs:2.95,0.8)[
  anchor=base west,
  text=black,
  rotate=0.0
]{ $\psi = 2.85$};
\end{axis}

\end{tikzpicture}}
		\subfloat{
\begin{tikzpicture}

\definecolor{color0}{rgb}{0.886274509803922,0.290196078431373,0.2}
\definecolor{color1}{rgb}{0.203921568627451,0.541176470588235,0.741176470588235}
\definecolor{color2}{rgb}{0.596078431372549,0.556862745098039,0.835294117647059}
\definecolor{color3}{rgb}{0.984313725490196,0.756862745098039,0.368627450980392}
\definecolor{color4}{rgb}{0.556862745098039,0.729411764705882,0.258823529411765}
\definecolor{color5}{rgb}{0.75,0,0.75}

\begin{axis}[
axis background/.style={fill=white},
axis line style={white!89.80392156862746!black},
legend cell align={left},
legend entries={\hspace{-.2cm}\textbf{$\rho$},{0.95},{0.88},{0.81},{0.72},{0.61},{0.38}},
legend style={draw=white!80.0!black, fill=white},
tick align=outside,
tick pos=left,
x grid style={white!89.80392156862746!black},
xlabel={Score $s$},
xmajorgrids,
xmin=0.8, xmax=5.2,
y grid style={white!89.80392156862746!black},
ylabel={$P(U = s)$},
ymajorgrids,
ymin=-0.05, ymax=1.05
]
\addlegendimage{empty legend}
\addplot [line width=0.08000000000000002pt, color0, dashed, mark=*, mark size=4, mark options={solid}]
table [row sep=\\]{%
1	0.00125295660335689 \\
2	0.0132129969081272 \\
3	0.130343269655477 \\
4	0.794662643551237 \\
5	0.0605281332818022 \\
};
\addplot [line width=0.08000000000000002pt, color1, dashed, mark=triangle*, mark size=4, mark options={solid,rotate=180}]
table [row sep=\\]{%
1	0.00300709584805654 \\
2	0.0317111925795053 \\
3	0.172823847173145 \\
4	0.647190344522968 \\
5	0.145267519876325 \\
};
\addplot [line width=0.08000000000000002pt, color2, dashed, mark=diamond*, mark size=4, mark options={solid}]
table [row sep=\\]{%
1	0.00476123509275618 \\
2	0.0502093882508834 \\
3	0.215304424690813 \\
4	0.4997180454947 \\
5	0.230006906470848 \\
};
\addplot [line width=0.08000000000000002pt, white!46.666666666666664!black, dashed, mark=x, mark size=4, mark options={solid}]
table [row sep=\\]{%
1	0.0131541005509078 \\
2	0.077735654200599 \\
3	0.221868434092756 \\
4	0.370439767009059 \\
5	0.316802044146677 \\
};
\addplot [line width=0.08000000000000002pt, color3, dashed, mark=pentagon*, mark size=4, mark options={solid}]
table [row sep=\\]{%
1	0.0368432269876218 \\
2	0.0997985070460573 \\
3	0.184045116936097 \\
4	0.285141337039146 \\
5	0.394171811991077 \\
};
\addplot [line width=0.08000000000000002pt, color4, dashed, mark=square*, mark size=4, mark options={solid}]
table [row sep=\\]{%
1	0.111056915197833 \\
2	0.0956263175840493 \\
3	0.107779556060854 \\
4	0.153334274334812 \\
5	0.532202936822451 \\
};
\addplot [very thick, color5, forget plot]
table [row sep=\\]{%
3.9	0 \\
3.9	1 \\
};
\node at (axis cs:2.9,0.8)[
  anchor=base west,
  text=black,
  rotate=0.0
]{ $\psi = 3.90$};
\end{axis}

\end{tikzpicture}}
		\caption{GSD distributions of $U$ for $M=5$ and for various values of $\psi$ and $\rho$.} \label{fig:disExa}
	\end{center}
\end{figure*}
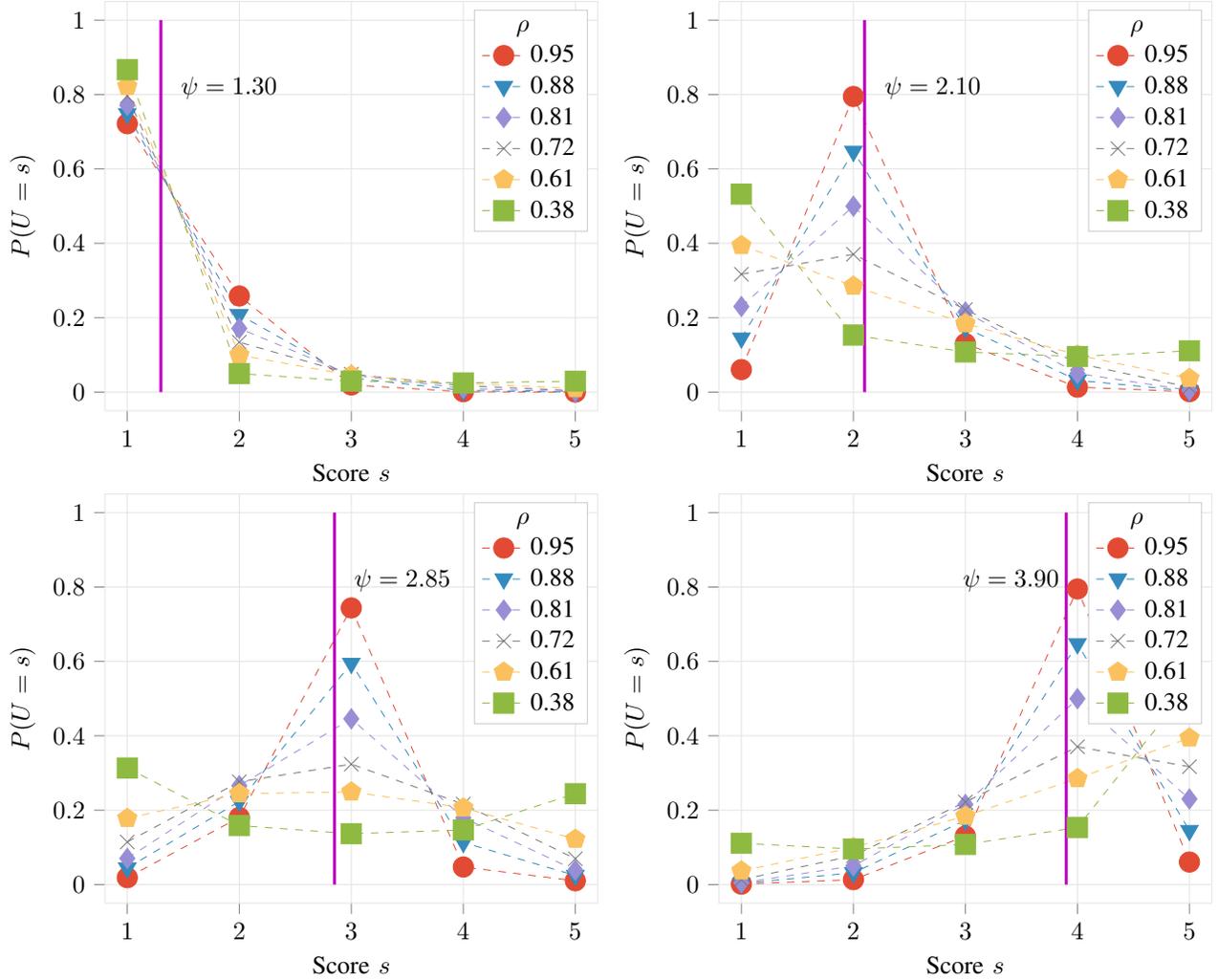

The motivation behind constructing the GSD class is to have a class that would properly describe response distributions observed when analysing responses from MQA subjective experiments. There, the responses are often expressed on a 5-level scale, with the following mapping between discrete consecutive numbers and textual labels: 1---Bad, 2---Poor, 3---Fair, 4---Good and 5---Excellent. Although initially designed for the MQA research, we expect the GSD class to be useful for describing other, more general processes (at least for measurement processes exhibiting a characteristic similar to what is the case for the MQA subjective experiments). Examples of specific incarnations of the GSD family of distributions (for $M=5$) are shown in Fig.~\ref{fig:disExa}. Note that for $\psi$ close to $1$ or $M$, regardless of $\rho$, the obtained distributions are similar. This is because the maximum spread we can obtain is limited by a small range of possible variances (see Fig.~\ref{fig:varRel}).


\subsection{Comparing Ordered Probit's and GSD's Parametrisation}
\label{ssec:comparing_op_a_gsd_param}
\begin{figure}[!t]
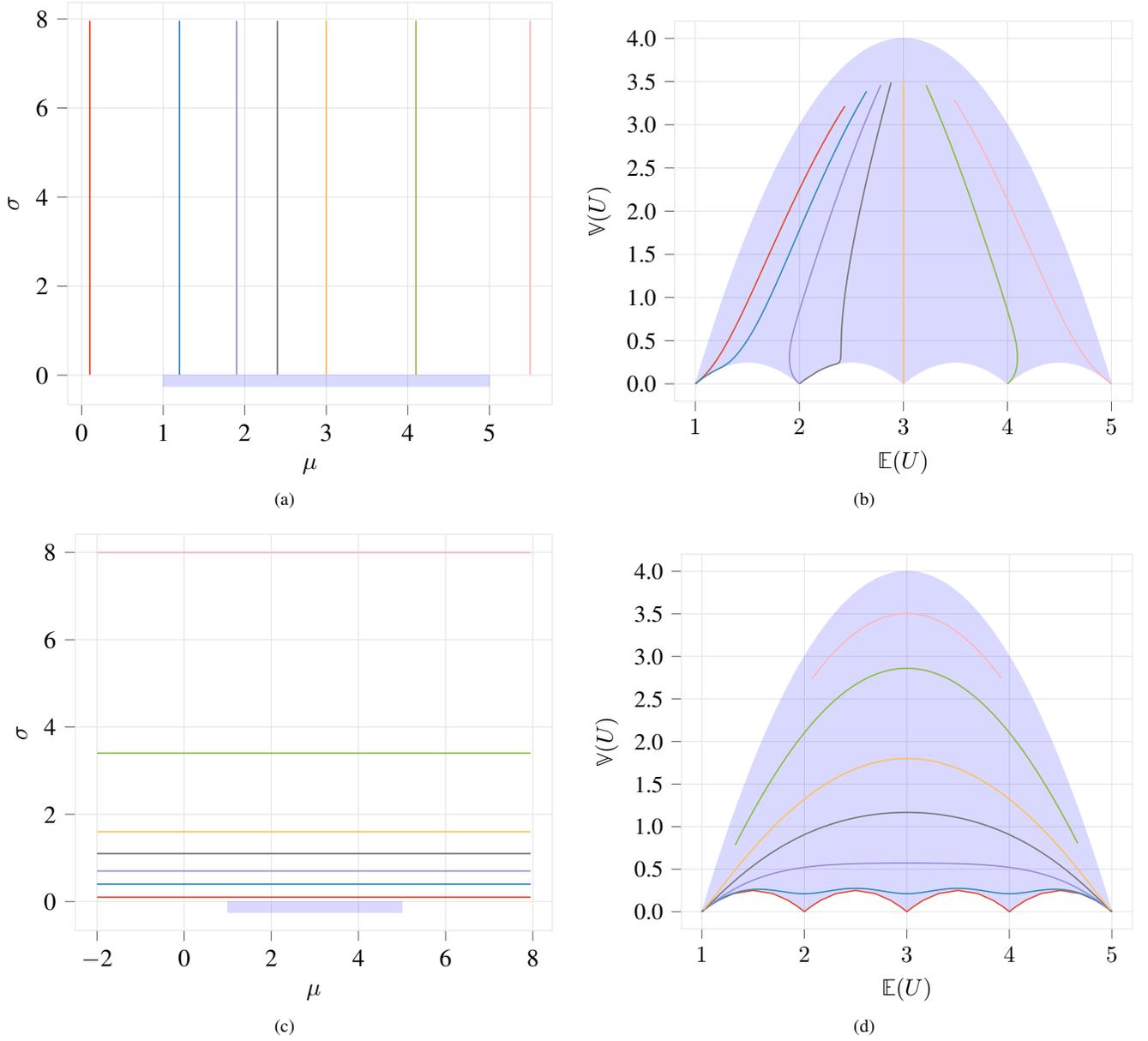

    \centering
    \subfloat[]{\resizebox{0.48\textwidth}{!}{\input{plots/space_mu.tex}}
        \label{fig:ghost_ordered_probit_a}}
    \hfil
    \subfloat[]{\resizebox{0.48\textwidth}{!}{\input{plots/ghost_mu.tex}}
        \label{fig:ghost_ordered_probit_b}}
    \\
    \subfloat[]{\resizebox{0.48\textwidth}{!}{
            \input{plots/space_sigma.tex}}
        \label{fig:ghost_ordered_probit_c}}
    \hfil
    \subfloat[]{\resizebox{0.48\textwidth}{!}{
            \input{plots/ghost_sigma.tex}}
        \label{fig:ghost_ordered_probit_d}}
    \caption{Mapping of Ordered Probit parameters to the $(\mathbb{E}(U), \mathbb{V}(U))$ space. The violet ghost-like area marks all possible ($\mathbb{E}(U), \mathbb{V}(U)$) pairs for a discrete process with values $\{1, 2, 3, 4, 5\}$.}
    \label{fig:ghost_ordered_probit}
\end{figure}

\begin{figure}[!t]
    \centering
    \subfloat[]{\resizebox{0.48\textwidth}{!}{
\begin{tikzpicture}

\definecolor{color0}{rgb}{0.886274509803922,0.290196078431373,0.2}
\definecolor{color1}{rgb}{0.203921568627451,0.541176470588235,0.741176470588235}
\definecolor{color2}{rgb}{0.596078431372549,0.556862745098039,0.835294117647059}
\definecolor{color3}{rgb}{0.984313725490196,0.756862745098039,0.368627450980392}
\definecolor{color4}{rgb}{0.556862745098039,0.729411764705882,0.258823529411765}
\definecolor{color5}{rgb}{1,0.709803921568627,0.72156862745098}

\begin{axis}[
axis background/.style={fill=white},
axis line style={white!89.8039215686275!black},
tick align=outside,
tick pos=left,
x grid style={white!89.8039215686275!black},
xlabel={\(\displaystyle \psi\)},
xmajorgrids,
xmin=0.8, xmax=5.2,
xtick style={color=white!33.3333333333333!black},
y grid style={white!89.8039215686275!black},
ylabel={\(\displaystyle \rho\)},
ymajorgrids,
ymin=-0.0828125, ymax=1.0515625,
ytick style={color=white!33.3333333333333!black},
ytick={-0.2,0,0.2,0.4,0.6,0.8,1,1.2},
yticklabels={\ensuremath{-}0.2,0.0,0.2,0.4,0.6,0.8,1.0,1.2}
]
\path [draw=blue, fill=blue, opacity=0.15, very thin]
(axis cs:1,0)
--(axis cs:1,-0.03125)
--(axis cs:5,-0.03125)
--(axis cs:5,0)
--(axis cs:5,0)
--(axis cs:1,0)
--cycle;

\addplot [semithick, color0]
table {%
1.1 0
1.1 0.05
1.1 0.1
1.1 0.15
1.1 0.2
1.1 0.25
1.1 0.3
1.1 0.35
1.1 0.4
1.1 0.45
1.1 0.5
1.1 0.55
1.1 0.6
1.1 0.65
1.1 0.7
1.1 0.75
1.1 0.8
1.1 0.85
1.1 0.9
1.1 0.95
1.1 1
};
\addplot [semithick, color1]
table {%
1.4 0
1.4 0.05
1.4 0.1
1.4 0.15
1.4 0.2
1.4 0.25
1.4 0.3
1.4 0.35
1.4 0.4
1.4 0.45
1.4 0.5
1.4 0.55
1.4 0.6
1.4 0.65
1.4 0.7
1.4 0.75
1.4 0.8
1.4 0.85
1.4 0.9
1.4 0.95
1.4 1
};
\addplot [semithick, color2]
table {%
2.1 0
2.1 0.05
2.1 0.1
2.1 0.15
2.1 0.2
2.1 0.25
2.1 0.3
2.1 0.35
2.1 0.4
2.1 0.45
2.1 0.5
2.1 0.55
2.1 0.6
2.1 0.65
2.1 0.7
2.1 0.75
2.1 0.8
2.1 0.85
2.1 0.9
2.1 0.95
2.1 1
};
\addplot [semithick, white!46.6666666666667!black]
table {%
2.6 0
2.6 0.05
2.6 0.1
2.6 0.15
2.6 0.2
2.6 0.25
2.6 0.3
2.6 0.35
2.6 0.4
2.6 0.45
2.6 0.5
2.6 0.55
2.6 0.6
2.6 0.65
2.6 0.7
2.6 0.75
2.6 0.8
2.6 0.85
2.6 0.9
2.6 0.95
2.6 1
};
\addplot [semithick, color3]
table {%
3 0
3 0.05
3 0.1
3 0.15
3 0.2
3 0.25
3 0.3
3 0.35
3 0.4
3 0.45
3 0.5
3 0.55
3 0.6
3 0.65
3 0.7
3 0.75
3 0.8
3 0.85
3 0.9
3 0.95
3 1
};
\addplot [semithick, color4]
table {%
3.9 0
3.9 0.05
3.9 0.1
3.9 0.15
3.9 0.2
3.9 0.25
3.9 0.3
3.9 0.35
3.9 0.4
3.9 0.45
3.9 0.5
3.9 0.55
3.9 0.6
3.9 0.65
3.9 0.7
3.9 0.75
3.9 0.8
3.9 0.85
3.9 0.9
3.9 0.95
3.9 1
};
\addplot [semithick, color5]
table {%
4.7 0
4.7 0.05
4.7 0.1
4.7 0.15
4.7 0.2
4.7 0.25
4.7 0.3
4.7 0.35
4.7 0.4
4.7 0.45
4.7 0.5
4.7 0.55
4.7 0.6
4.7 0.65
4.7 0.7
4.7 0.75
4.7 0.8
4.7 0.85
4.7 0.9
4.7 0.95
4.7 1
};
\end{axis}

\end{tikzpicture}}
        \label{fig:ghost_gsd_a}}
    \hfil
    \subfloat[]{\resizebox{0.48\textwidth}{!}{
\begin{tikzpicture}

\definecolor{color0}{rgb}{0.886274509803922,0.290196078431373,0.2}
\definecolor{color1}{rgb}{0.203921568627451,0.541176470588235,0.741176470588235}
\definecolor{color2}{rgb}{0.596078431372549,0.556862745098039,0.835294117647059}
\definecolor{color3}{rgb}{0.984313725490196,0.756862745098039,0.368627450980392}
\definecolor{color4}{rgb}{0.556862745098039,0.729411764705882,0.258823529411765}
\definecolor{color5}{rgb}{1,0.709803921568627,0.72156862745098}

\begin{axis}[
axis background/.style={fill=white},
axis line style={white!89.8039215686275!black},
tick align=outside,
tick pos=left,
x grid style={white!89.8039215686275!black},
xlabel={\(\displaystyle \mathbb{E}(U)\)},
xmajorgrids,
xmin=0.8, xmax=5.2,
xtick style={color=white!33.3333333333333!black},
y grid style={white!89.8039215686275!black},
ylabel={\(\displaystyle \mathbb{V}(U)\)},
ymajorgrids,
ymin=-0.200000000000015, ymax=4.2,
ytick style={color=white!33.3333333333333!black},
ytick={-0.5,0,0.5,1,1.5,2,2.5,3,3.5,4,4.5},
yticklabels={\ensuremath{-}0.5,0.0,0.5,1.0,1.5,2.0,2.5,3.0,3.5,4.0,4.5}
]
\path [draw=blue, fill=blue, opacity=0.15, very thin]
(axis cs:1,0)
--(axis cs:1,0)
--(axis cs:1.05,0.0475)
--(axis cs:1.1,0.0900000000000001)
--(axis cs:1.15,0.1275)
--(axis cs:1.2,0.16)
--(axis cs:1.25,0.1875)
--(axis cs:1.3,0.21)
--(axis cs:1.35,0.2275)
--(axis cs:1.4,0.24)
--(axis cs:1.45,0.2475)
--(axis cs:1.5,0.25)
--(axis cs:1.55,0.2475)
--(axis cs:1.6,0.24)
--(axis cs:1.65,0.2275)
--(axis cs:1.7,0.21)
--(axis cs:1.75,0.1875)
--(axis cs:1.8,0.16)
--(axis cs:1.85,0.127499999999999)
--(axis cs:1.9,0.0899999999999994)
--(axis cs:1.95,0.0474999999999992)
--(axis cs:2,8.88178419700124e-16)
--(axis cs:2.05,0.0475000000000006)
--(axis cs:2.1,0.0900000000000008)
--(axis cs:2.15,0.127500000000001)
--(axis cs:2.2,0.160000000000001)
--(axis cs:2.25,0.1875)
--(axis cs:2.3,0.21)
--(axis cs:2.35,0.2275)
--(axis cs:2.4,0.24)
--(axis cs:2.45,0.2475)
--(axis cs:2.5,0.25)
--(axis cs:2.55,0.2475)
--(axis cs:2.6,0.24)
--(axis cs:2.65,0.2275)
--(axis cs:2.7,0.209999999999999)
--(axis cs:2.75,0.187499999999999)
--(axis cs:2.8,0.159999999999999)
--(axis cs:2.85,0.127499999999999)
--(axis cs:2.9,0.0899999999999987)
--(axis cs:2.95,0.0474999999999982)
--(axis cs:3,1.77635683940025e-15)
--(axis cs:3.05,0.0475000000000014)
--(axis cs:3.1,0.0900000000000015)
--(axis cs:3.15,0.127500000000002)
--(axis cs:3.2,0.160000000000001)
--(axis cs:3.25,0.187500000000001)
--(axis cs:3.3,0.210000000000001)
--(axis cs:3.35,0.227500000000001)
--(axis cs:3.4,0.24)
--(axis cs:3.45,0.2475)
--(axis cs:3.5,0.25)
--(axis cs:3.55,0.2475)
--(axis cs:3.6,0.24)
--(axis cs:3.65,0.227499999999999)
--(axis cs:3.7,0.209999999999999)
--(axis cs:3.75,0.187499999999999)
--(axis cs:3.8,0.159999999999999)
--(axis cs:3.85,0.127499999999998)
--(axis cs:3.9,0.0899999999999979)
--(axis cs:3.95,0.0474999999999974)
--(axis cs:4,2.66453525910037e-15)
--(axis cs:4.05,0.0475000000000022)
--(axis cs:4.1,0.0900000000000026)
--(axis cs:4.15,0.127500000000002)
--(axis cs:4.2,0.160000000000002)
--(axis cs:4.25,0.187500000000001)
--(axis cs:4.3,0.210000000000001)
--(axis cs:4.35,0.227500000000001)
--(axis cs:4.4,0.240000000000001)
--(axis cs:4.45,0.2475)
--(axis cs:4.5,0.25)
--(axis cs:4.55,0.2475)
--(axis cs:4.6,0.239999999999999)
--(axis cs:4.65,0.227499999999999)
--(axis cs:4.7,0.209999999999999)
--(axis cs:4.75,0.187499999999998)
--(axis cs:4.8,0.159999999999998)
--(axis cs:4.85,0.127499999999998)
--(axis cs:4.9,0.0899999999999969)
--(axis cs:4.95,0.0474999999999966)
--(axis cs:5,3.55271367880049e-15)
--(axis cs:5,-1.4210854715202e-14)
--(axis cs:5,-1.4210854715202e-14)
--(axis cs:4.95,0.197499999999985)
--(axis cs:4.9,0.389999999999985)
--(axis cs:4.85,0.577499999999988)
--(axis cs:4.8,0.759999999999988)
--(axis cs:4.75,0.937499999999988)
--(axis cs:4.7,1.10999999999999)
--(axis cs:4.65,1.27749999999999)
--(axis cs:4.6,1.43999999999999)
--(axis cs:4.55,1.59749999999999)
--(axis cs:4.5,1.74999999999999)
--(axis cs:4.45,1.89749999999999)
--(axis cs:4.4,2.03999999999999)
--(axis cs:4.35,2.17749999999999)
--(axis cs:4.3,2.30999999999999)
--(axis cs:4.25,2.43749999999999)
--(axis cs:4.2,2.55999999999999)
--(axis cs:4.15,2.67749999999999)
--(axis cs:4.1,2.78999999999999)
--(axis cs:4.05,2.89749999999999)
--(axis cs:4,2.99999999999999)
--(axis cs:3.95,3.09749999999999)
--(axis cs:3.9,3.19)
--(axis cs:3.85,3.2775)
--(axis cs:3.8,3.36)
--(axis cs:3.75,3.4375)
--(axis cs:3.7,3.51)
--(axis cs:3.65,3.5775)
--(axis cs:3.6,3.64)
--(axis cs:3.55,3.6975)
--(axis cs:3.5,3.75)
--(axis cs:3.45,3.7975)
--(axis cs:3.4,3.84)
--(axis cs:3.35,3.8775)
--(axis cs:3.3,3.91)
--(axis cs:3.25,3.9375)
--(axis cs:3.2,3.96)
--(axis cs:3.15,3.9775)
--(axis cs:3.1,3.99)
--(axis cs:3.05,3.9975)
--(axis cs:3,4)
--(axis cs:2.95,3.9975)
--(axis cs:2.9,3.99)
--(axis cs:2.85,3.9775)
--(axis cs:2.8,3.96)
--(axis cs:2.75,3.9375)
--(axis cs:2.7,3.91)
--(axis cs:2.65,3.8775)
--(axis cs:2.6,3.84)
--(axis cs:2.55,3.7975)
--(axis cs:2.5,3.75)
--(axis cs:2.45,3.6975)
--(axis cs:2.4,3.64)
--(axis cs:2.35,3.5775)
--(axis cs:2.3,3.51)
--(axis cs:2.25,3.4375)
--(axis cs:2.2,3.36)
--(axis cs:2.15,3.2775)
--(axis cs:2.1,3.19)
--(axis cs:2.05,3.0975)
--(axis cs:2,3)
--(axis cs:1.95,2.8975)
--(axis cs:1.9,2.79)
--(axis cs:1.85,2.6775)
--(axis cs:1.8,2.56)
--(axis cs:1.75,2.4375)
--(axis cs:1.7,2.31)
--(axis cs:1.65,2.1775)
--(axis cs:1.6,2.04)
--(axis cs:1.55,1.8975)
--(axis cs:1.5,1.75)
--(axis cs:1.45,1.5975)
--(axis cs:1.4,1.44)
--(axis cs:1.35,1.2775)
--(axis cs:1.3,1.11)
--(axis cs:1.25,0.937500000000001)
--(axis cs:1.2,0.760000000000001)
--(axis cs:1.15,0.5775)
--(axis cs:1.1,0.39)
--(axis cs:1.05,0.1975)
--(axis cs:1,0)
--cycle;

\addplot [semithick, color0]
table {%
1.1 0.39
1.1 0.375
1.1 0.36
1.1 0.345
1.1 0.33
1.1 0.315
1.1 0.3
1.1 0.285
1.1 0.27
1.1 0.255
1.1 0.24
1.1 0.225
1.1 0.21
1.1 0.195
1.1 0.18
1.1 0.165
1.1 0.15
1.1 0.135
1.1 0.12
1.1 0.105
1.1 0.0900000000000001
};
\addplot [semithick, color1]
table {%
1.4 1.44
1.4 1.38
1.4 1.32
1.4 1.26
1.4 1.2
1.4 1.14
1.4 1.08
1.4 1.02
1.4 0.96
1.4 0.9
1.4 0.84
1.4 0.78
1.4 0.72
1.4 0.66
1.4 0.6
1.4 0.54
1.4 0.48
1.4 0.42
1.4 0.36
1.4 0.3
1.4 0.24
};
\addplot [semithick, color2]
table {%
2.1 3.19
2.1 3.035
2.1 2.88
2.1 2.725
2.1 2.57
2.1 2.415
2.1 2.26
2.1 2.105
2.1 1.95
2.1 1.795
2.1 1.64
2.1 1.485
2.1 1.33
2.1 1.175
2.1 1.02
2.1 0.865000000000001
2.1 0.71
2.1 0.555000000000001
2.1 0.4
2.1 0.244999999999999
2.1 0.0899999999999999
};
\addplot [semithick, white!46.6666666666667!black]
table {%
2.6 3.84
2.6 3.66
2.6 3.48
2.6 3.3
2.6 3.12
2.6 2.94
2.6 2.76
2.6 2.58
2.6 2.4
2.6 2.22
2.6 2.04
2.6 1.86
2.6 1.68
2.6 1.5
2.6 1.32
2.6 1.14
2.6 0.959999999999999
2.6 0.779999999999999
2.6 0.599999999999999
2.6 0.419999999999999
2.6 0.239999999999999
};
\addplot [semithick, color3]
table {%
3 4
3 3.8
3 3.6
3 3.4
3 3.2
3 3
3 2.8
3 2.6
3 2.4
3 2.2
3 2
3 1.8
3 1.6
3 1.4
3 1.2
3 1
3 0.800000000000001
3 0.6
3 0.4
3 0.199999999999996
3 0
};
\addplot [semithick, color4]
table {%
3.9 3.19
3.9 3.035
3.9 2.88
3.9 2.725
3.9 2.57
3.9 2.415
3.9 2.26
3.9 2.105
3.9 1.95
3.9 1.795
3.9 1.64
3.9 1.485
3.9 1.33
3.9 1.175
3.9 1.02
3.9 0.865000000000004
3.9 0.709999999999997
3.9 0.554999999999998
3.9 0.4
3.9 0.245000000000001
3.9 0.0899999999999999
};
\addplot [semithick, color5]
table {%
4.7 1.11
4.7 1.065
4.7 1.02
4.7 0.975000000000001
4.7 0.929999999999993
4.7 0.884999999999994
4.7 0.839999999999996
4.7 0.794999999999998
4.7 0.749999999999993
4.7 0.704999999999995
4.7 0.659999999999997
4.7 0.614999999999998
4.7 0.569999999999993
4.7 0.524999999999995
4.7 0.480000000000004
4.7 0.434999999999999
4.7 0.389999999999997
4.7 0.345000000000006
4.7 0.300000000000004
4.7 0.254999999999995
4.7 0.209999999999997
};
\end{axis}

\end{tikzpicture}}
        \label{fig:ghost_gsd_b}}
    \\
    \subfloat[]{\resizebox{0.48\textwidth}{!}{
\begin{tikzpicture}

\definecolor{color0}{rgb}{0.886274509803922,0.290196078431373,0.2}
\definecolor{color1}{rgb}{0.203921568627451,0.541176470588235,0.741176470588235}
\definecolor{color2}{rgb}{0.596078431372549,0.556862745098039,0.835294117647059}
\definecolor{color3}{rgb}{0.984313725490196,0.756862745098039,0.368627450980392}
\definecolor{color4}{rgb}{0.556862745098039,0.729411764705882,0.258823529411765}
\definecolor{color5}{rgb}{1,0.709803921568627,0.72156862745098}

\begin{axis}[
axis background/.style={fill=white},
axis line style={white!89.8039215686275!black},
tick align=outside,
tick pos=left,
x grid style={white!89.8039215686275!black},
xlabel={\(\displaystyle \psi\)},
xmajorgrids,
xmin=0.8, xmax=5.2,
xtick style={color=white!33.3333333333333!black},
y grid style={white!89.8039215686275!black},
ylabel={\(\displaystyle \rho\)},
ymajorgrids,
ymin=-0.0828125, ymax=1.0515625,
ytick style={color=white!33.3333333333333!black},
ytick={-0.2,0,0.2,0.4,0.6,0.8,1,1.2},
yticklabels={\ensuremath{-}0.2,0.0,0.2,0.4,0.6,0.8,1.0,1.2}
]
\path [draw=blue, fill=blue, opacity=0.15, very thin]
(axis cs:1,0)
--(axis cs:1,-0.03125)
--(axis cs:5,-0.03125)
--(axis cs:5,0)
--(axis cs:5,0)
--(axis cs:1,0)
--cycle;

\addplot [semithick, color0]
table {%
1.01 0.01
1.06 0.01
1.11 0.01
1.16 0.01
1.21 0.01
1.26 0.01
1.31 0.01
1.36 0.01
1.41 0.01
1.46 0.01
1.51 0.01
1.56 0.01
1.61 0.01
1.66 0.01
1.71 0.01
1.76 0.01
1.81 0.01
1.86 0.01
1.91 0.01
1.96 0.01
2.01 0.01
2.06 0.01
2.11 0.01
2.16 0.01
2.21 0.01
2.26 0.01
2.31 0.01
2.36 0.01
2.41 0.01
2.46 0.01
2.51 0.01
2.56 0.01
2.61 0.01
2.66 0.01
2.71 0.01
2.76 0.01
2.81 0.01
2.86 0.01
2.91 0.01
2.96 0.01
3.01 0.01
3.06 0.01
3.11 0.01
3.16 0.01
3.21 0.01
3.26 0.01
3.31 0.01
3.36 0.01
3.41 0.01
3.46 0.01
3.51 0.01
3.56 0.01
3.61 0.01
3.66 0.01
3.71 0.01
3.76 0.01
3.81 0.01
3.86 0.01
3.91 0.01
3.96 0.01
4.01 0.01
4.06 0.01
4.11 0.01
4.16 0.01
4.21 0.01
4.26 0.01
4.31 0.01
4.36 0.01
4.41 0.01
4.46 0.01
4.51 0.01
4.56 0.01
4.61 0.01
4.66 0.01
4.71 0.01
4.76 0.01
4.81 0.01
4.86 0.01
4.91 0.01
4.96 0.01
4.99 0.01
};
\addplot [semithick, color1]
table {%
1.01 0.1
1.06 0.1
1.11 0.1
1.16 0.1
1.21 0.1
1.26 0.1
1.31 0.1
1.36 0.1
1.41 0.1
1.46 0.1
1.51 0.1
1.56 0.1
1.61 0.1
1.66 0.1
1.71 0.1
1.76 0.1
1.81 0.1
1.86 0.1
1.91 0.1
1.96 0.1
2.01 0.1
2.06 0.1
2.11 0.1
2.16 0.1
2.21 0.1
2.26 0.1
2.31 0.1
2.36 0.1
2.41 0.1
2.46 0.1
2.51 0.1
2.56 0.1
2.61 0.1
2.66 0.1
2.71 0.1
2.76 0.1
2.81 0.1
2.86 0.1
2.91 0.1
2.96 0.1
3.01 0.1
3.06 0.1
3.11 0.1
3.16 0.1
3.21 0.1
3.26 0.1
3.31 0.1
3.36 0.1
3.41 0.1
3.46 0.1
3.51 0.1
3.56 0.1
3.61 0.1
3.66 0.1
3.71 0.1
3.76 0.1
3.81 0.1
3.86 0.1
3.91 0.1
3.96 0.1
4.01 0.1
4.06 0.1
4.11 0.1
4.16 0.1
4.21 0.1
4.26 0.1
4.31 0.1
4.36 0.1
4.41 0.1
4.46 0.1
4.51 0.1
4.56 0.1
4.61 0.1
4.66 0.1
4.71 0.1
4.76 0.1
4.81 0.1
4.86 0.1
4.91 0.1
4.96 0.1
4.99 0.1
};
\addplot [semithick, color2]
table {%
1.01 0.4
1.06 0.4
1.11 0.4
1.16 0.4
1.21 0.4
1.26 0.4
1.31 0.4
1.36 0.4
1.41 0.4
1.46 0.4
1.51 0.4
1.56 0.4
1.61 0.4
1.66 0.4
1.71 0.4
1.76 0.4
1.81 0.4
1.86 0.4
1.91 0.4
1.96 0.4
2.01 0.4
2.06 0.4
2.11 0.4
2.16 0.4
2.21 0.4
2.26 0.4
2.31 0.4
2.36 0.4
2.41 0.4
2.46 0.4
2.51 0.4
2.56 0.4
2.61 0.4
2.66 0.4
2.71 0.4
2.76 0.4
2.81 0.4
2.86 0.4
2.91 0.4
2.96 0.4
3.01 0.4
3.06 0.4
3.11 0.4
3.16 0.4
3.21 0.4
3.26 0.4
3.31 0.4
3.36 0.4
3.41 0.4
3.46 0.4
3.51 0.4
3.56 0.4
3.61 0.4
3.66 0.4
3.71 0.4
3.76 0.4
3.81 0.4
3.86 0.4
3.91 0.4
3.96 0.4
4.01 0.4
4.06 0.4
4.11 0.4
4.16 0.4
4.21 0.4
4.26 0.4
4.31 0.4
4.36 0.4
4.41 0.4
4.46 0.4
4.51 0.4
4.56 0.4
4.61 0.4
4.66 0.4
4.71 0.4
4.76 0.4
4.81 0.4
4.86 0.4
4.91 0.4
4.96 0.4
4.99 0.4
};
\addplot [semithick, white!46.6666666666667!black]
table {%
1.01 0.6
1.06 0.6
1.11 0.6
1.16 0.6
1.21 0.6
1.26 0.6
1.31 0.6
1.36 0.6
1.41 0.6
1.46 0.6
1.51 0.6
1.56 0.6
1.61 0.6
1.66 0.6
1.71 0.6
1.76 0.6
1.81 0.6
1.86 0.6
1.91 0.6
1.96 0.6
2.01 0.6
2.06 0.6
2.11 0.6
2.16 0.6
2.21 0.6
2.26 0.6
2.31 0.6
2.36 0.6
2.41 0.6
2.46 0.6
2.51 0.6
2.56 0.6
2.61 0.6
2.66 0.6
2.71 0.6
2.76 0.6
2.81 0.6
2.86 0.6
2.91 0.6
2.96 0.6
3.01 0.6
3.06 0.6
3.11 0.6
3.16 0.6
3.21 0.6
3.26 0.6
3.31 0.6
3.36 0.6
3.41 0.6
3.46 0.6
3.51 0.6
3.56 0.6
3.61 0.6
3.66 0.6
3.71 0.6
3.76 0.6
3.81 0.6
3.86 0.6
3.91 0.6
3.96 0.6
4.01 0.6
4.06 0.6
4.11 0.6
4.16 0.6
4.21 0.6
4.26 0.6
4.31 0.6
4.36 0.6
4.41 0.6
4.46 0.6
4.51 0.6
4.56 0.6
4.61 0.6
4.66 0.6
4.71 0.6
4.76 0.6
4.81 0.6
4.86 0.6
4.91 0.6
4.96 0.6
4.99 0.6
};
\addplot [semithick, color3]
table {%
1.01 0.85
1.06 0.85
1.11 0.85
1.16 0.85
1.21 0.85
1.26 0.85
1.31 0.85
1.36 0.85
1.41 0.85
1.46 0.85
1.51 0.85
1.56 0.85
1.61 0.85
1.66 0.85
1.71 0.85
1.76 0.85
1.81 0.85
1.86 0.85
1.91 0.85
1.96 0.85
2.01 0.85
2.06 0.85
2.11 0.85
2.16 0.85
2.21 0.85
2.26 0.85
2.31 0.85
2.36 0.85
2.41 0.85
2.46 0.85
2.51 0.85
2.56 0.85
2.61 0.85
2.66 0.85
2.71 0.85
2.76 0.85
2.81 0.85
2.86 0.85
2.91 0.85
2.96 0.85
3.01 0.85
3.06 0.85
3.11 0.85
3.16 0.85
3.21 0.85
3.26 0.85
3.31 0.85
3.36 0.85
3.41 0.85
3.46 0.85
3.51 0.85
3.56 0.85
3.61 0.85
3.66 0.85
3.71 0.85
3.76 0.85
3.81 0.85
3.86 0.85
3.91 0.85
3.96 0.85
4.01 0.85
4.06 0.85
4.11 0.85
4.16 0.85
4.21 0.85
4.26 0.85
4.31 0.85
4.36 0.85
4.41 0.85
4.46 0.85
4.51 0.85
4.56 0.85
4.61 0.85
4.66 0.85
4.71 0.85
4.76 0.85
4.81 0.85
4.86 0.85
4.91 0.85
4.96 0.85
4.99 0.85
};
\addplot [semithick, color4]
table {%
1.01 0.92
1.06 0.92
1.11 0.92
1.16 0.92
1.21 0.92
1.26 0.92
1.31 0.92
1.36 0.92
1.41 0.92
1.46 0.92
1.51 0.92
1.56 0.92
1.61 0.92
1.66 0.92
1.71 0.92
1.76 0.92
1.81 0.92
1.86 0.92
1.91 0.92
1.96 0.92
2.01 0.92
2.06 0.92
2.11 0.92
2.16 0.92
2.21 0.92
2.26 0.92
2.31 0.92
2.36 0.92
2.41 0.92
2.46 0.92
2.51 0.92
2.56 0.92
2.61 0.92
2.66 0.92
2.71 0.92
2.76 0.92
2.81 0.92
2.86 0.92
2.91 0.92
2.96 0.92
3.01 0.92
3.06 0.92
3.11 0.92
3.16 0.92
3.21 0.92
3.26 0.92
3.31 0.92
3.36 0.92
3.41 0.92
3.46 0.92
3.51 0.92
3.56 0.92
3.61 0.92
3.66 0.92
3.71 0.92
3.76 0.92
3.81 0.92
3.86 0.92
3.91 0.92
3.96 0.92
4.01 0.92
4.06 0.92
4.11 0.92
4.16 0.92
4.21 0.92
4.26 0.92
4.31 0.92
4.36 0.92
4.41 0.92
4.46 0.92
4.51 0.92
4.56 0.92
4.61 0.92
4.66 0.92
4.71 0.92
4.76 0.92
4.81 0.92
4.86 0.92
4.91 0.92
4.96 0.92
4.99 0.92
};
\addplot [semithick, color5]
table {%
1.01 1
1.06 1
1.11 1
1.16 1
1.21 1
1.26 1
1.31 1
1.36 1
1.41 1
1.46 1
1.51 1
1.56 1
1.61 1
1.66 1
1.71 1
1.76 1
1.81 1
1.86 1
1.91 1
1.96 1
2.01 1
2.06 1
2.11 1
2.16 1
2.21 1
2.26 1
2.31 1
2.36 1
2.41 1
2.46 1
2.51 1
2.56 1
2.61 1
2.66 1
2.71 1
2.76 1
2.81 1
2.86 1
2.91 1
2.96 1
3.01 1
3.06 1
3.11 1
3.16 1
3.21 1
3.26 1
3.31 1
3.36 1
3.41 1
3.46 1
3.51 1
3.56 1
3.61 1
3.66 1
3.71 1
3.76 1
3.81 1
3.86 1
3.91 1
3.96 1
4.01 1
4.06 1
4.11 1
4.16 1
4.21 1
4.26 1
4.31 1
4.36 1
4.41 1
4.46 1
4.51 1
4.56 1
4.61 1
4.66 1
4.71 1
4.76 1
4.81 1
4.86 1
4.91 1
4.96 1
4.99 1
};
\end{axis}

\end{tikzpicture}}
        \label{fig:ghost_gsd_c}}
    \hfil
    \subfloat[]{\resizebox{0.48\textwidth}{!}{
            \input{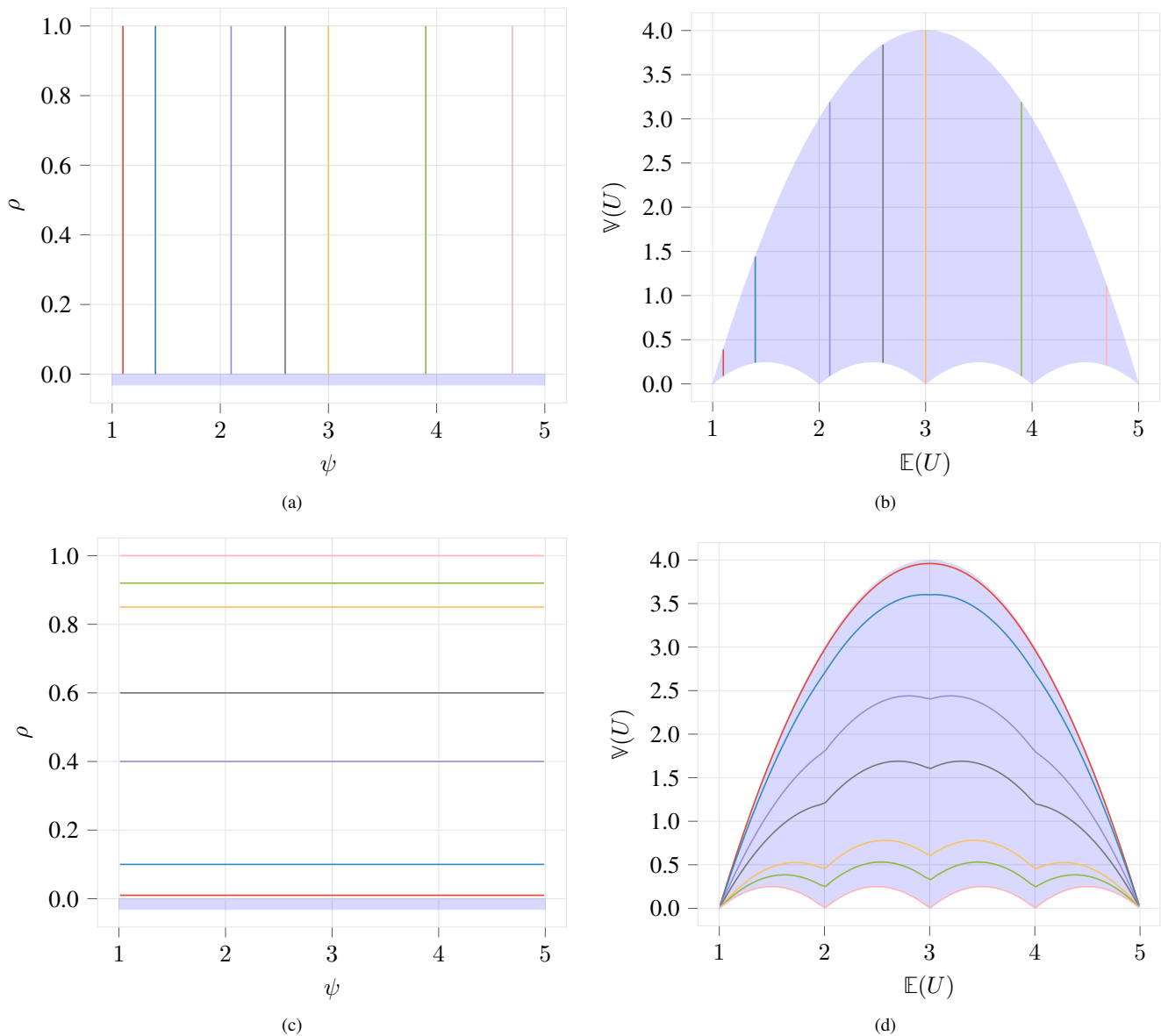}}
        \label{fig:ghost_gsd_d}}
    \caption{Mapping of GSD parameters to the $(\mathbb{E}(U), \mathbb{V}(U))$ space. The violet ghost-like area marks all possible ($\mathbb{E}(U), \mathbb{V}(U)$) pairs for a discrete process with values $\{1, 2, 3, 4, 5\}$.}
    \label{fig:ghost_gsd}
\end{figure}

In this section, we present the interaction between ordered probit parameters, GSD parameters, and the $(\mathbb{E}(U), \mathbb{V}(U))$ space. Fig.~\ref{fig:ghost_ordered_probit} presents the interaction between ordered probit parameters ($\mu$ and $\sigma$) and summary statistics ($\mathbb{E}(U)$ and $\mathbb{V}(U)$). The latter are calculated taking discrete responses generated by the ordered probit model with a given $\mu$ and $\sigma$ pair. The lines present in the left-hand-side of Fig.~\ref{fig:ghost_ordered_probit}, correspond to the same coloured lines in the right-hand-side of the same figure. Specifically, the ordering of lines (when going from left to right in Fig.~\ref{fig:ghost_ordered_probit_a} and top to bottom in Fig.~\ref{fig:ghost_ordered_probit_c}) in the left-hand-side of the figure is the same as the ordering of lines in the right-hand-side of the figure.
Fig.~\ref{fig:ghost_gsd} presents the corresponding plots for the GSD class. Note, however, that the ordering of lines in Fig.~\ref{fig:ghost_gsd_c} is reversed to the ordering of lines in Fig.~\ref{fig:ghost_gsd_d}. Differently put, the top-most line in Fig.~\ref{fig:ghost_gsd_c} corresponds to the bottom-most line in Fig.~\ref{fig:ghost_gsd_d}. Importantly, both Fig.~\ref{fig:ghost_ordered_probit} and Fig.~\ref{fig:ghost_gsd} use $M=5$.

Fig.~\ref{fig:ghost_ordered_probit} is a graphical presentation of the problems inherent to the estimation and interpretation of ordered probit parameters, when these are based on the observations of the random variable $U$.
First of all, the mapping from any bounded set of $(\mu,\sigma)$ pairs, results in a set of $(\mathbb{E}(U),\mathbb{V}(U))$ pairs that does not cover the whole ghost-like area of all possible $(\mathbb{E}(U),\mathbb{V}(U))$ pairs. Notice that the lines in Fig.~\ref{fig:ghost_ordered_probit_b} do not reach neither the sides nor the top of the ghost-like area.
Second of all, there is no analytical formula mapping $(\mathbb{E}(U),\mathbb{V}(U))$ pairs to $(\mu,\sigma)$ pairs. It is difficult to even approximately guess which $(\mu,\sigma)$ pair corresponds to which $(\mathbb{E}(U),\mathbb{V}(U))$ pair.
This is a big limitation of ordered probit's parameterisation. The estimation of a $(\mathbb{E}(U),\mathbb{V}(U))$ pair, based on observations $U_1,...,U_n$, is an easy task. Unfortunately, the results of this estimation are not useful for estimating ordered probit parameters.

The problems described above do not apply to GSD's parameterisation. The $\psi$ parameter is the expected value of the observations ($U_1,...,U_n$). Thus, for any $(\mathbb{E}(U),\mathbb{V}(U))$ pair, we immediately know the corresponding $\psi$. This is because $\psi$ is equal to $\mathbb{E}(U)$. Notice that the vertical lines in Fig. \ref{fig:ghost_gsd_a} and Fig. \ref{fig:ghost_gsd_b} are in identical positions along the horizontal axis.
The second GSD class' parameter, $\rho$, is the confidence parameter. Its value of $0$ corresponds to the biggest possible variance (the upper bound of the violet coloured ghost-like area in Fig. \ref{fig:ghost_gsd_b}) and $1$ corresponds to the smallest possible variance (the lower bound of the violet coloured ghost-like area in Fig. \ref{fig:ghost_gsd_b}) of the observations.
Moreover, the range of $\rho$ values is linear. For example, $\rho=0.7$ can be interpreted as $70\%$, in terms of the available variance. That is, $\mathbb{V}(U)=70\% V_{\mathrm{min}}(\psi)+30\%V_{\mathrm{max}}(\psi)$. Therefore, to obtain $\rho$ from a $(\mathbb{E}(U),\mathbb{V}(U))$ pair, it is enough to calculate the distance between $\mathbb{V}(U)$ and the upper bound of the violet coloured ghost-like area in Fig. \ref{fig:ghost_gsd_b}. This distance should be then divided by the distance between the upper and lower bounds of the violet coloured ghost-like area in Fig. \ref{fig:ghost_gsd_b}. Differently put, 
$\rho=\frac{V_{\mathrm{max}}(\psi)-\mathbb{V}(U)}{V_{\mathrm{max}}(\psi)-V_{\mathrm{min}}(\psi)}$. Value of $\rho$ is thus simply a ratio between the distance between the observed variance ($\mathbb{V}(U)$) and the maximum possible variance ($V_{\mathrm{max}}$) and the available range of variance ($V_{\mathrm{max}}(\psi)-V_{\mathrm{min}}(\psi)$).
The easy mapping of $(\mathbb{E}(U),\mathbb{V}(U))$ pairs to $(\psi,\rho)$ pairs makes the estimation of $(\psi,\rho)$ pairs (when based on $U_1,...,U_n$ observations) much easier than the estimation of $(\mu,\sigma)$ pairs for the ordered probit model. 

We would also like to draw the reader's attention to the violet horizontal bars present in Fig.~\ref{fig:ghost_ordered_probit_a}, \ref{fig:ghost_ordered_probit_c}, \ref{fig:ghost_gsd_a}, and \ref{fig:ghost_gsd_c}. In all cases, the violet bars span the range from 1 to 5. This range corresponds to the width of the violet ghost-like area in the right-hand side of Fig.~\ref{fig:ghost_ordered_probit} and ~\ref{fig:ghost_gsd}. Please note that GSD's parameterisation is bounded to this 1 to 5 range, whereas ordered probit's one is not. This is yet another advantageous feature of GSD's parameterisation. Being bounded to the same range as the range of observable responses, GSD class parameters are easier to interpret and understand.

\section{GSD Parameters Estimation}
\label{sec:estimation}

Ordered probit and GSD models cannot be used without an accurate and efficient parameter estimation procedure. The simplest approach is to use the method of moments. As we mentioned in the previous section, the method of moments for the ordered probit model is rather problematic. However, for the GSD class, moments based estimation is quite simple, i.e.,
$$(\hat{\psi},\hat{\rho})=\left(\widehat{\mathbb{E}(U)},\frac{V_{\mathrm{max}}(\widehat{\mathbb{E}(U)})-\widehat{\mathbb{V}(U)}}{V_{\mathrm{max}}(\widehat{\mathbb{E}(U)})-V_{\mathrm{min}}(\widehat{\mathbb{E}(U)})}\right),$$
where $\widehat{\mathbb{E}(U)}$, $\widehat{\mathbb{V}(U)}$ are expectation value and variance of the empirical distribution. To compare the GSD with ordered probit and to apply the likelihood ratio test of goodness-of-fit, we actually use Maximum Likelihood Estimator (MLE) for both models.
To make the comparison between the two models fair, when fitting them to real data, we use the same numerical estimation method for both, i.e., we use the estimation using a dense grid of points. Specifically, we first compute probabilities of all response categories for a set of $(\psi, \rho)$ (or $(\mu, \sigma)$) pairs and then search through the resultant grid to find the pair best matching the sample of interest.
In the multidimensional case (cf. Appendix~\ref{app:multi}), for generated data, we use the gradient based estimation method for the GSD class. The moments based estimator serves as a starting point. The exact formulae for the log-likelihood function and gradient that were used in the estimation algorithm are in Appendix \ref{sec:ap:est}.


\subsection{Numerical Experiments for the GSD Class}
\label{ssec:num_exp_for_gsd_class}
To validate our MLE procedure, we perform a simulation study. We draw data from the proposed distribution and then estimate the distribution parameters using the samples generated. Importantly, we do so for the case of $M=5$.  

\begin{figure}
	\begin{center}
        \includegraphics[scale=0.7]{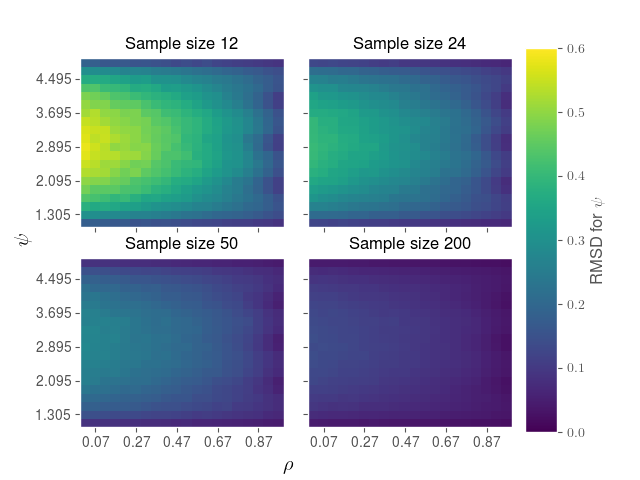}
		\caption{Root Mean Square Distance (RMSD) between the input $\psi$ and the estimated $\hat{\psi}$ for different $(\psi, \rho)$, different sample sizes, and $M=5$. For every sample size we generated $500\,000$ samples to obtain the figure.} \label{fig:RMSD_psi}
	\end{center}
\end{figure}

In Fig.~\ref{fig:RMSD_psi} we present the risk measured as Root Mean Square Distance (RMSD) between true value $\psi$ and estimated $\hat{\psi}$, for sample sizes $n=12, 24, 50, 200$. As one can see, the hardest case is when $\rho$ is small and $\psi$ is in the middle of the $[1,M]$ scale.
This behaviour is expected, since in the middle of the scale, the possible variance is the largest (cf. Fig.~\ref{fig:ghost_ordered_probit_b}).
One can also see that the parameter $\psi$ is rather easy to estimate. That is, the risk is rather small for $\psi\in[1,5]$, even for small sample sizes. 

\begin{figure}
	\begin{center}
        \includegraphics[scale=0.7]{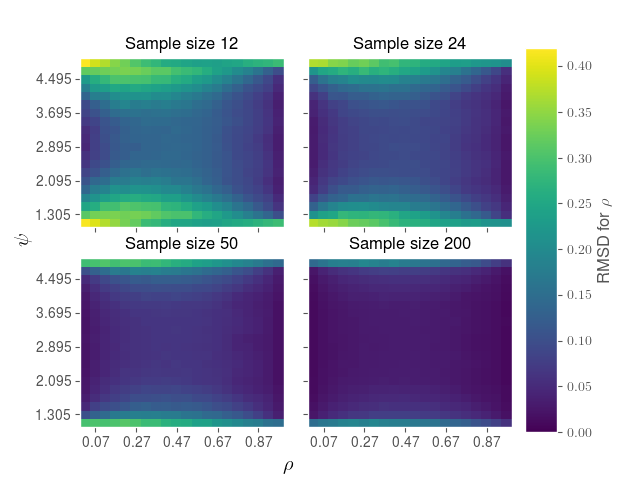}
		\caption{Root Mean Square Distance (RMSD) between the input $\rho$ and the estimated $\hat{\rho}$ for different $(\psi, \rho)$, different sample sizes, and $M=5$. For every sample size we generated $500\,000$ samples to obtain the figure. } \label{fig:RMSD_rho}
	\end{center}
\end{figure}

In Fig.~\ref{fig:RMSD_rho} we present the risk measured as the RMSD between true value $\rho$ and estimated $\hat{\rho}$, for sample sizes $n=12, 24, 50, 200$. In this case, the situation is slightly more complicated (than that presented above for $\psi$).
The hardest case is when $\psi$ is on the edge of the $[1,M]$ interval. As one can see in Fig.~\ref{fig:varMOS}, when $\psi$ is close to the edge of the $[1,M]$ interval, variance expressed as a function of $\rho$ is almost horizontal. This means that even small changes in sample variance correspond to large changes of $\rho$.
Simply put, the smaller the $[V_{\mathrm{min}}(\psi),V_{\mathrm{max}}(\psi)]$ interval, the harder the estimation of $\rho$. It is worth pointing out, however, that even for hard cases the risk is getting smaller, the larger is the sample size.

We also validate the MLE procedure for the multidimensional case. Specifically, we generate responses of $n$ raters (also referred to as subjects), each having assigned a confidence parameter $\rho_1,...,\rho_n$. The subjects rate $m$ objects (also referred to as stimuli), each having assigned an expectation value $\psi_1,...,\psi_m$ (which can be, for example, interpreted as latent true qualities of a set of $m$ videos presented to $n$ subjects). Based on the responses generated, we used the multidimensional MLE to recover the GSD parameters. More details and estimation results are in Appendix~\ref{app:multi}. 


\section{Real Data Examples}
\label{sec:analysis}
In the previous sections, we presented a new GSD family of distributions and showed that the estimation method properly extracts GSD parameters when applied to the simulated data. In this section, we validate if the proposed GSD class can be used to model real subjective data (i.e., subjective responses coming from MQA experiments).


\subsection{Comparing Goodness-of-Fit of Ordered Probit and GSD for Multimedia Quality Assessment Data}
\label{ssec:gof_op_vs_gsd}
We want to compare the GSD with the ordered probit model and one state-of-the-art solution. We come up with the latter by adapting the model presented in~\cite{Li2020a}. We call the adapted version of the model \textit{Simplified Li2020} or SLI for short.\footnote{For a detailed description regarding how did we transform the model from~\cite{Li2020a}, we refer the reader to Sec.~III-D of~\cite{Nawala2022}.}

To check whether a distribution fits specific data, we have to perform a two-step
procedure. The first step is to estimate distribution parameters for a sample of interest.
The second step is to test a null hypothesis, saying that the sample truly comes from
the assumed distribution (GSD, ordered probit or SLI), given the parameters estimated in
the first step. We choose a standard likelihood ratio approach to test the goodness-of-fit (GoF) of the models.
Specifically, we use the G-test of GoF (cf. Sec. 14.3.4 of~\cite{Agresti}). 
Since sample sizes we consider are mostly small (less than 30 observations per sample), we do not use the asymptotic
distribution for calculating the $p$-value. Conversely, we estimate the $p$-value
using a bootstrapped version of the G-test (see Appendix \ref{gtest}). (For comprehensive theoretical considerations
on the topic please take a look at~\cite{Efron}.)

\begin{figure}
	\centering
    \includegraphics[width=0.7\textwidth]{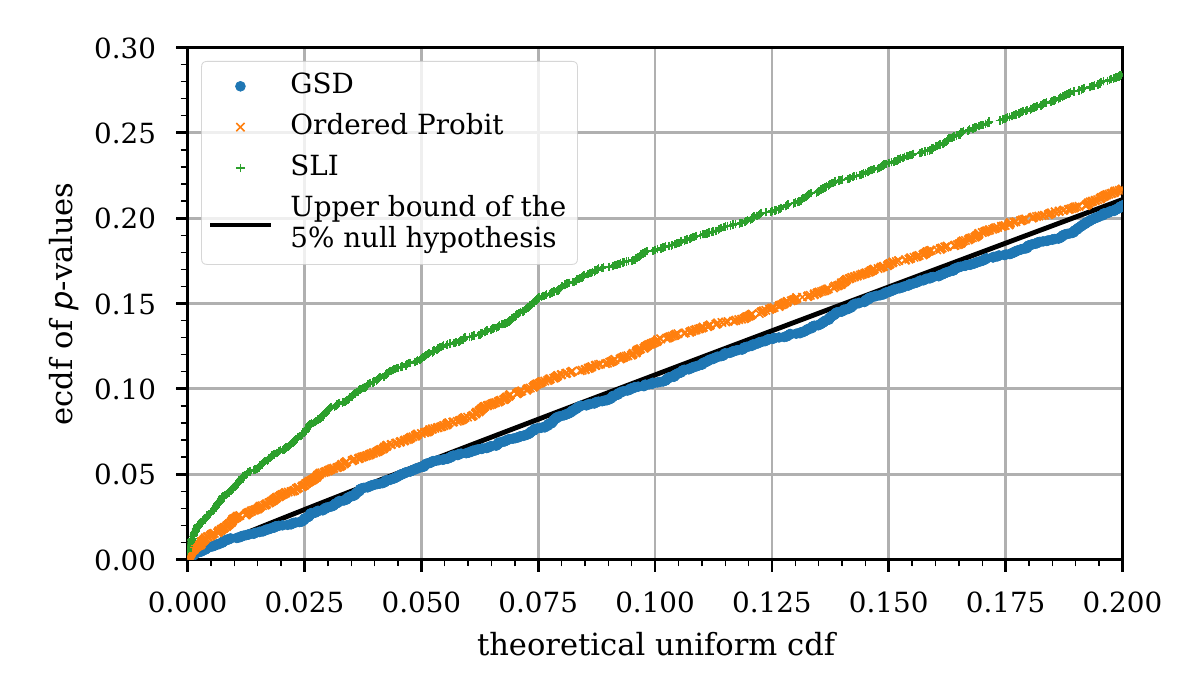}
	\caption{$p$-Value P–P plot for typical multimedia quality assessment (MQA) experiments. $p$-Values come from the G-test of goodness-of-fit applied to the GSD, ordered probit and Simplified Li2020 (SLI) models fitted to responses from 24 real-life subjective experiments. Ecdf stands for empirical cumulative distribution function.}
	\label{fig:p_value_pp_plot_mqa_exp}
\end{figure}
The data we use to perform the analysis come from six MQA studies: (i) ITU~\cite{itutsupp23},
(ii) HDTV~\cite{HDTV_Phase_I_test}, (iii) MM2~\cite{Pinson2012},
(iv) 14-505~\cite{AGH_NTIA_14-505}, (v) ITS4S~\cite{ITS4S} and
(vi) NFLX. We do not provide here extensive details regarding each study. Instead,
we refer the reader to publications cited next to each acronym. Since the NFLX study
does not have a dedicated publication, we refer the reader to Sec. II.C of~\cite{Nawala2022}.
Two important features of the six studies is that they all focus on Multimedia Quality
Assessment (MQA) and follow best practices and recommendations in the field. In other words,
we can safely call them \textit{typical} MQA studies. Some studies represent data from more
than one subjective experiment. Differently put, one study may consist of multiple experiments. In total, we use data from 24 subjective experiments. This
amounts to more than $100\,000$ individual scores (exactly $111\,198$) for more than
$3\,500$ stimuli (exactly $3\,643$).

In Fig.~\ref{fig:p_value_pp_plot_mqa_exp} we present the cumulative distribution function (CDF) of GoF test $p$-values for the GSD, ordered probit, and SLI models. The black line is the upper bound of $95\%$ right-sided confidence interval for the CDF of $p$-values under the null hypothesis. Specifically, under the null hypothesis, the CDF of $p$-values is not greater than the uniform distribution function (for more details see \cite{Nawala2020ACM}). As one can see, there is no evidence that the GSD is not the correct way of modelling subjective responses from MQA experiments. On the other hand, there is evidence that the distributions modelled by the ordered probit and SLI models are not suitable here.


\subsection{Bootstrapping}
\label{ssec:bootstrapping}
If we would like to use the bootstrap technique in some testing problems with data expressed on a Likert scale, there are two approaches to resampling that one can consider. One can either use the empirical distribution or fit a distribution coming from some assumed parametric class.
Here, we show that for at least one type of real data, specifically responses coming from MQA experiments, it is better to use the estimated GSD than it is to use the empirical distribution. This holds at least in the case of relatively small sample sizes.
(In the field of MQA, usually only up to 30 responses per stimulus are available.) To compare the behaviour of empirical probability mass function (EPMF) and the GSD, we use the following algorithm.

Let us denote by $N$ the number of observations in the large sample 
(e.g., $N = 200$) and
by $n$ the number of observations in the subsample of this large sample
(e.g., $n = 24$).
Now, we denote by $\left( N_1, N_2, N_3, N_4, N_5 \right)$ the
frequencies of each response category in the large sample. We denote
by $\left( p_1, p_2, p_3, p_4, p_5 \right)$ the EPMF
of the large sample. The test procedure is as follows (assuming there are five response categories):
\begin{enumerate}
    \item Generate $MC$ bootstrap samples (e.g., $MC = 10\,000$) of size $n$
    from the EPMF of the large
    sample $\left( p_1, p_2, p_3, p_4, p_5 \right)$.
    \item For the $r$-th bootstrap sample
    ($r = 1, 2, \ldots, MC$) do the following.
    \begin{enumerate}
        \item Estimate response category
        probabilities using maximum likelihood estimation for the model
        of interest (e.g., the GSD model). Denote the estimated
        probabilities by $\left( \hat{q}_1, \hat{q}_2, \hat{q}_3, \hat{q_4},
        \hat{q}_5 \right)$.
        \item Denote by $\left( \hat{v}_1, \hat{v}_2, \hat{v}_3, \hat{v}_4,
        \hat{v}_5 \right)$ the
        EPMF of the bootstrap sample.
        \item Find the likelihood $\mathcal{L}_m$ of the estimated model
        for the large sample.
        In other words, calculate 
        $$
        \mathcal{L}_m = \prod\limits_{k=1, N_k \neq 0}^{5} \hat{q}_k^{N_k}.
        $$
        \item Find the likelihood $\mathcal{L}_e$ of the
        bootstrap sample's EPMF
        for the large sample. In other words, calculate
        $$
        \mathcal{L}_e = \prod\limits_{k=1, N_k \neq 0}^{5} \hat{v}_k^{N_k}.
        $$ 
        \item Find the natural logarithm of the ratio of the two likelihoods
        and denote it by $W_r$
        $$
        W_r = \ln \left( \frac{\mathcal{L}_m}{\mathcal{L}_e}  \right).
        $$
        Note that the above simplifies to
        $$
        W_r = \sum\limits_{k=1, N_k \neq 0}^{5} N_k \left( \ln \hat{q}_k - \ln
        \hat{v}_k \right).
        $$
    \end{enumerate}
    \item Calculate the estimator of $p_{\mathrm{GSD}}-p_{\mathrm{e}}=P(W_r > 0)-P(W_r < 0)$, which is the difference between the probability that the GSD has greater likelihood than the
    EPMF and the probability that the EPMF has greater likelihood than the GSD.
    This can be formally described by the following.
    $$
    \hat{p}_{\mathrm{GSD}}-\hat{p}_{\mathrm{e}}= \frac{\sum\limits_{r=1}^{MC} I \left( W_r > 0 \right)}{MC}-\frac{\sum\limits_{r=1}^{MC}I \left( W_r < 0 \right)}{MC},
    $$
    where $I(x)$ is one if $x$ is true or 0 if $x$ is false.
    \item Calculate $.95$ confidence interval for $p_{\mathrm{GSD}}-p_{\mathrm{e}}$ i.e.,
    $$L=\hat{p}_{\mathrm{GSD}}-\hat{p}_{\mathrm{e}}-1.96\sqrt{\frac{\hat{p}_{\mathrm{GSD}}+\hat{p}_{\mathrm{e}}-(\hat{p}_{\mathrm{GSD}}-\hat{p}_{\mathrm{e}})^2}{MC}}$$
    $$R=\hat{p}_{\mathrm{GSD}}-\hat{p}_{\mathrm{e}}+1.96\sqrt{\frac{\hat{p}_{\mathrm{GSD}}+\hat{p}_{\mathrm{e}}-(\hat{p}_{\mathrm{GSD}}-\hat{p}_{\mathrm{e}})^2}{MC}}$$
    For $L>0$ the GSD performs better. For $R<0$ the EPMF performs better.
    If $[L,R]$ contains zero there is no significant difference between
    the GSD and EPMF.
\end{enumerate}

We use data from four MQA studies: (i) MM2~\cite{Pinson2012}, (ii) HDTV~\cite{HDTV_Phase_I_test}, (iii) NFLX (cf. Sec. II.C of~\cite{Nawala2022}) and (iv) ITERO~\cite{Perez2021}. We describe the first three studies in Sec.~\ref{ssec:gof_op_vs_gsd}. The last study, contrary to the first three, is not a typical MQA study. We decide to use it anyway for two reasons. First, being atypical, it should not not give unfair advantage to the GSD. Second, it provides real data with many responses per stimulus. This last property also stands behind our choice to use the other three studies (i.e., MM2, HDTV and NFLX). Specifically, we only select from these stimuli with at least 144 responses. This results in 234 stimuli, each assigned between 144 and 228 responses.

We use three small sample sizes, i.e., $n = \{ 12, 24, 50 \}$.
This way we can observe how the GSD performs (when compared to the
empirical distribution) for different fractions of the large sample
information available. Intuitively, we expect the empirical distribution's
performance to improve as the small sample size increases. If the
GSD proofs to perform differently than the empirical distribution
we would observe how the increasing small sample size influences
the difference between the two approaches. We emphasise here that
the increasing small sample size always favours the empirical distribution.
On the other hand, the performance of the GSD depends on how well
it fits to the distribution of responses observed in the large sample.
If the fit is good, increasing small sample size also favours
the GSD. If the fit is poor, increasing small sample size does
not necessarily improve GSD's performance.

\begin{figure}[!t]
	\centering
	\includegraphics[width=.45\textwidth]{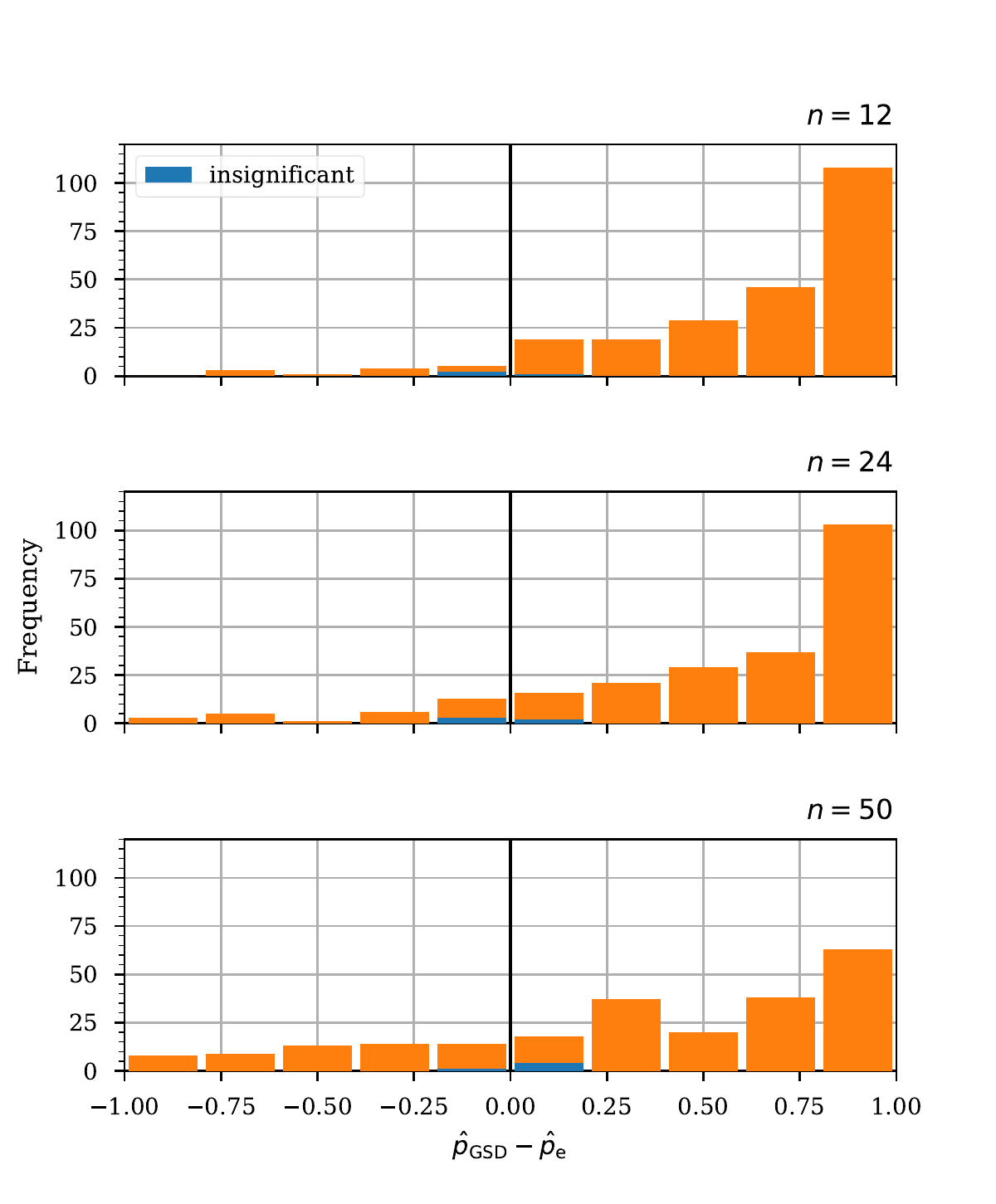}
	\caption{Histograms depicting the distribution of probability differences
	$\hat{p}_{\mathrm{GSD}}-\hat{p}_{\mathrm{e}}$ for three different small
	sample sizes (i.e., 12, 24 and 50) and for the case of the (unmodified)
	GSD being compared with the (unmodified) empirical distribution.
	Blue-coloured parts of the bars represent statistically
	insignificant probability differences.}
	\label{fig:prob_diff_histograms_unmodified}
\end{figure}

Fig.~\ref{fig:prob_diff_histograms_unmodified} presents results
of the analysis. It contains three histograms of probability differences
$\hat{p}_{\mathrm{GSD}}-\hat{p}_{\mathrm{e}}$. Each histogram shows results for one of the investigated small sample sizes (i.e., 12, 24, and 50). Larger probability mass to the right of zero means the GSD outperforms the empirical distribution. Larger probability mass to the left of zero means the empirical distribution performs better than the GSD. As can be seen, the GSD outperforms the empirical distribution for all three small sample sizes considered.

\begin{figure}[!t]
	\centering
	\includegraphics[width=.45\textwidth]{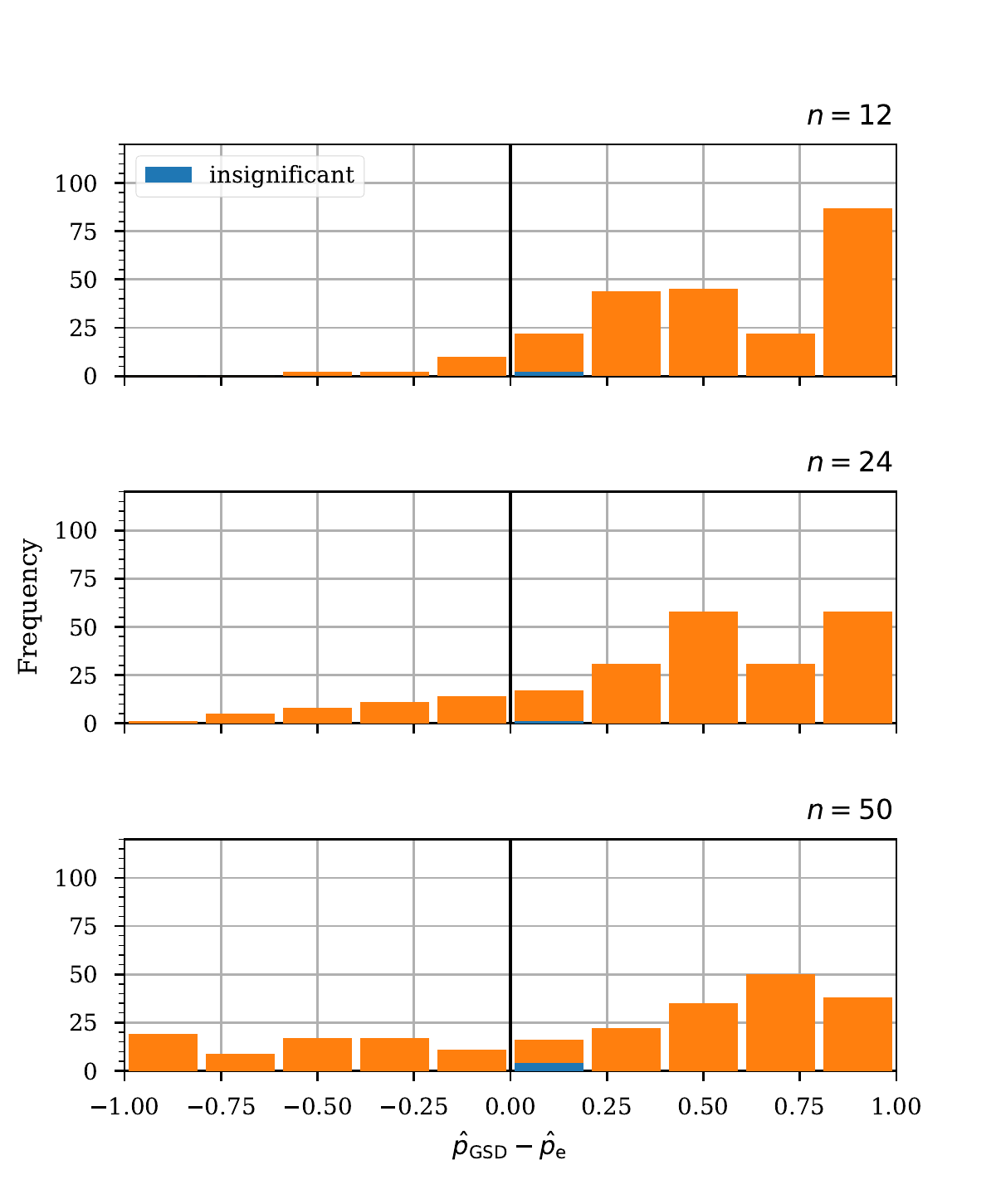}
	\caption{Histograms depicting the distribution of probability differences
	$\hat{p}_{\mathrm{GSD}}-\hat{p}_{\mathrm{e}}$ for three different small
	sample sizes (i.e., 12, 24 and 50) and for the case of the corrected GSD
	being compared with the corrected empirical distribution. Blue-coloured
	parts of the bars represent statistically insignificant
	probability differences.}
	\label{fig:prob_diff_histograms_corrected}
\end{figure}

We theorise that the reason the GSD outperforms the empirical distribution is because the latter assigns a larger than the GSD probability to empty cells (i.e., response categories with no responses assigned to them in the sample of interest). To verify this claim, we run our analysis once again, this time modifying the estimation procedure both for the GSD and empirical distribution. This modification does not allow any empty cells. In other words, the estimated probability of any response category has to be necessarily in the interval $(0,1)$. The details are in Appendix \ref{app:parameters_estimation_modification}. Fig.~\ref{fig:prob_diff_histograms_corrected} presents results
for the case of the corrected GSD being compared with the corrected
empirical distribution. Again, the GSD outperforms the empirical distribution, although this time by a smaller margin.

The results clearly show that the GSD is a better choice than the empirical distribution
when it comes to resampling of subjective responses from MQA studies. 


\section{Conclusion}
In this paper, we propose a Generalised Score Distribution (GSD) class. It is a family of discrete distributions with: finite support, two parameters, and no more than one change in probability monotonicity. The distribution parameters are: $\psi$ determining the mean, and $\rho$ determining the spread of the responses. 

We show the usefulness of the GSD class for modelling, with a special focus on the Multimedia Quality Assessment (MQA) field. The class is a convenient regularisation of the multinomial distribution. The GSD class has only two parameters and covers all possible first- and second-order moments for a distribution defined on a discrete finite support. We also evidence that the GSD class can be useful in testing problems using the parametric bootstrap technique.

The advantage of the GSD class is that its $\rho$ parameter can be used to determine the type of the underlining process. With $\rho$ close to 1, we know that the process is similar to the Bernoulli distribution. Likewise, for $\rho < C(\psi)$ we know the process rather resembles the beta-binomial distribution. This information can be used as a diagnostic tool, answering the following question: ``What is the spread of the responses?'' Note that the GSD class can be easily used outside of the MQA field, wherever information about responses spread is relevant.

\bctodo{Wspomniałeś kiedyś ``Trzeba powołać się na dane o jakości mierzonej liczbowo''. Czy pamiętasz o co Ci chodziło?}
We strongly believe that the GSD class can be of use for modelling results similar in nature to those reported in the field of MQA. More specifically, the GSD can be potentially useful for modelling subjective responses, where the population of observers generally agree about a given trait of a stimulus presented to them (e.g., about the visual quality of a distorted image). To put this differently, the GSD will not likely work in cases where there are evident subgroups of observers. For example, when there are two groups that have opposing views on a stimulus trait, they are asked to assess. The only exception to GSD's inability of modelling opposing views is the so-called ``love or hate'' case. In this case, a significant proportion of the population of observers either scores a stimulus trait extremely high or extremely low (cf. Fig.~\ref{fig:disExa}, the GSD distribution with $\psi=2.85$ and $\rho=0.38$).

In the future research, we would like to add more parameters to the GSD class. One idea is to add a parameter related to a potential personal bias of each observer (similarly to what is done in \cite{Janowski2015} and \cite{Li2017}). This subject bias parameter may, for example, show whether a person is generally more optimistic than other raters are. Another interesting direction of research would be using the GSD for data other than that coming from MQA subjective experiments. We would like to collaborate with scientists working in different fields, from audio and image quality, through student performance assessment, and up to psychology and sociology. In all those fields, a proper modelling of the response generation process would help to gain new insights. Our results obtained for MQA subjective data might be treated as a proof of concept, showing that the GSD class may be of use for those other fields as well.


\section*{Acknowledgements}
The authors would like to thank Netflix, Inc. for sponsoring the initial phases of this research. This work was also partially supported by the PL-Grid Infrastructure.
We warmly acknowledge comments and suggestions from Zhi Li of Netflix as well. We would also like to thank Anush Krishna Moorthy for coordinating the NFLX experiment and providing valuable feedback. Furthermore, we would like to thank Pablo Pérez, Narciso García and Margaret Pinson, for creating the ITERO data set, which helped us perform the bootstrap analysis (cf.~Sec.~\ref{ssec:bootstrapping}). 


\bibliographystyle{IEEEtran}
%

\bibliography{bibliography}
\appendices


\section{Proofs}\label{proofs}
Proof of Proposition \ref{GSDbetabinomial}:

First notice that $P_{G_\rho}(\epsilon=k-\psi)$ (see formula (\ref{eq:Prho})) can be rewritten as
$$P_{G_\rho}(\epsilon=k-\psi)=\binom{M-1}{k-1}
 \frac{\mathcal{B}\left(\frac{(\psi-1)\rho}{(M-1)(C(\psi)-\rho)}+k-1,\frac{(M-\psi)\rho}{(M-1)(C(\psi)-\rho)}+M-k\right)}{\mathcal{B}\left(\frac{(\psi-1)\rho}{(M-1)(C(\psi)-\rho)},\frac{(M-\psi)\rho}{(M-1)(C(\psi)-\rho)}\right)}$$
 Now observe that $\epsilon+\psi-1$ has the beta-binomial distribution $BB(M-1,\alpha,\beta)$ with parameters 
 $$\alpha=\frac{(\psi-1)\rho}{(M-1)(C(\psi)-\rho)}, \ \ \ \beta=\frac{(M-\psi)\rho}{(M-1)(C(\psi)-\rho)}.$$
 Using formula for beta-binomial expectation value we obtain
 $$\mathbb{E}(U)=\mathbb{E}(\psi+\epsilon)=\frac{(M-1)\alpha}{\alpha+\beta}+1=\psi.$$
  Using formula for beta-binomial variance we obtain
  $$\mathbb{V}(U)=\mathbb{V}(\psi+\epsilon)=\mathbb{V}(\psi+\epsilon-1)=\frac{(M-1)\alpha\beta(\alpha+\beta+M-1)}{(\alpha+\beta)^2(\alpha+\beta+1)}=\frac{(\psi-1)(M-\psi)}{M-1}\left(1+\frac{M-2}{C(\psi)}(C(\psi)-\rho)\right).$$
  Since $(\psi-1)(M-\psi)=V_{\mathrm{max}}(\psi)$ and $C(\psi)=\frac{M-2}{M-1}\ \frac{V_{\mathrm{max}}(\psi)}{V_{\mathrm{max}}(\psi)-V_{\mathrm{min}}(\psi)}$ (see formulas (\ref{Varmin}), (\ref{Varmax}) and (\ref{Cpsi})) we have
$$\mathbb{V}(U)=\frac{V_{\mathrm{max}}(\psi)}{M-1}+(V_{\mathrm{max}}(\psi)-V_{\mathrm{min}}(\psi))\left(\frac{M-2}{M-1} \frac{V_{\mathrm{max}}(\psi)}{V_{\mathrm{max}}(\psi)-V_{\mathrm{min}}(\psi)}-\rho\right)=\rho V_{\mathrm{min}}(\psi)+(1-\rho)V_{\mathrm{max}}(\psi)$$
\\
\\
Proof of Proposition \ref{GSDmix}:

In $F_{\rho}$ distribution case (see formula (\ref{eq:Frho})) notice that the random variable $\epsilon$ is a mixture of random variables $\epsilon_1$ and $\epsilon_2$, i.e.,
$$\epsilon=D\epsilon_1+(1-D)\epsilon_2,$$
where
$$P(\epsilon_1=k-\psi)= [1-|k-\psi|]_{+}, \ \ \ P(\epsilon_2=k-\psi)= 
		 \binom{M-1}{k-1}\left(\frac{\psi-1}{M-1}\right)^{k-1}\left(\frac{M-\psi}{M-1}\right)^{M-k},$$
$$P(D=1)=\frac{\rho-C(\psi)}{1-C(\psi)}, \ \ \ P(D=0)=\frac{1-\rho}{1-C(\psi)},$$
and $\epsilon_1, \epsilon_2, D$ are independent.
If $\psi$ is an integer $P(\epsilon_1=0)=1$ so $\mathbb{E}(\epsilon_1)=0$. In case $\psi$ is not an integer we have
$$\mathbb{E}(\epsilon_1)=(1+\lfloor\psi\rfloor-\psi)(\lfloor\psi\rfloor-\psi)+(\psi-\lfloor\psi\rfloor)(1+\lfloor\psi\rfloor-\psi)=0.$$
In both cases $\mathbb{E}(\epsilon_1)=0$. Now, observe that $\epsilon_2+\psi-1$ has the binomial distribution $B(M-1,p)$ with $p=\frac{\psi-1}{M-1}$. Therefore
$$\mathbb{E}(\epsilon_2)=(M-1)\frac{\psi-1}{M-1}-\psi+1=0.$$
We have then
$$\mathbb{E}(U)=\mathbb{E}(\epsilon+\psi)=\mathbb{E}(\epsilon_1+\psi|D=1)P(D=1)+\mathbb{E}(\epsilon_2+\psi|D=0)P(D=0)=\psi$$
Now, notice that $\mathbb{V}(\epsilon_1)=V_{\mathrm{min}}(\psi)$ and $V(\epsilon_2)=V_{\mathrm{Bin}}(\psi)$ (see formulas (\ref{Varmin}), (\ref{Varmax}) and (\ref{eq:VarBin})). Therefore
$$\mathbb{V}(U)=\mathbb{V}(\epsilon)=\mathbb{E}(\epsilon^2)=\mathbb{E}(\epsilon_1^2)\frac{\rho-C(\psi)}{1-C(\psi)}+\mathbb{E}(\epsilon_2^2)\frac{1-\rho}{1-C(\psi)}=V_{\mathrm{min}}(\psi)\frac{\rho-C(\psi)}{1-C(\psi)}+V_{\mathrm{Bin}}(\psi)\frac{1-\rho}{1-C(\psi)}=:f_{\psi}(\rho).$$
We want to show that for every fixed $\psi\in [1,M]$, the variance $\mathbb{V}(U)$ is the linear function of $\rho$ equal to $\rho V_{\mathrm{min}}(\psi)+(1-\rho)V_{\mathrm{max}}(\psi)$. Notice that the $f_\psi$ function is a linear function of variable $\rho$. Therefore, it is enough to show that 
$$\frac{d}{d\rho}f_\psi(\rho)=\frac{d}{d\rho}(\rho V_{\mathrm{min}}(\psi)+(1-\rho)V_{\mathrm{max}}(\psi))=V_{\mathrm{min}}(\psi)-V_{\mathrm{max}}(\psi)$$
and $f_\psi(1) = V_{\mathrm{min}}(\psi)$. It is easy to see that $f_\psi(1) = V_{\mathrm{min}}(\psi)$ so let us check the derivative. 
Since $C(\psi)=\frac{M-2}{M-1}\ \frac{V_{\mathrm{max}}(\psi)}{V_{\mathrm{max}}(\psi)-V_{\mathrm{min}}(\psi)}$ and $V_{\mathrm{Bin}}(\psi)=\frac{V_{\mathrm{max}}(\psi)}{M-1}$, we obtain
$$\frac{d}{d\rho}f_\psi(\rho)=\frac{(V_{\mathrm{max}}(\psi)-V_{\mathrm{min}}(\psi))((M-1)V_{\mathrm{min}}(\psi)-V_{\mathrm{max}}(\psi))}{(M-1)(V_{\mathrm{max}}(\psi)-V_{\mathrm{min}}(\psi))-(M-2)V_{\mathrm{max}}(\psi)}=V_{\mathrm{min}}(\psi)-V_{\mathrm{max}}(\psi)$$
Therefore
$$\mathbb{V}(U)=\rho V_{\mathrm{min}}(\psi)+(1-\rho)V_{\mathrm{max}}(\psi).$$
\\
\\
Proof of Proposition \ref{GSDber}:\\
First notice that in the case of $\rho\geq C(\psi)$
$$
P(Z_i=1)=\begin{cases} \frac{\rho-C(\psi)}{1-C(\psi)}+\frac{1-\rho}{1-C(\psi)}\frac{\psi-1}{M-1} \ \ \mathrm{for}\ \ i\leq \lfloor\psi\rfloor-1\\
\frac{\rho-C(\psi)}{1-C(\psi)}(\psi+1-\lceil\psi \rceil)+\frac{1-\rho}{1-C(\psi)}\frac{\psi-1}{M-1} \ \ \mathrm{for}\ \ i=\lceil\psi \rceil-1\\ 
\frac{1-\rho}{1-C(\psi)}\frac{\psi-1}{M-1} \ \ \mathrm{for}\ \ i\geq\lceil\psi \rceil \end{cases}.
$$
Random variables $Z_i$ can be written as $Z_i=DX_i+(1-D)Y_i$ where $X_i,Y_i,D$ are independent zero-one random variables and
$$P(D=1)=\frac{\rho-C(\psi)}{1-C(\psi)}$$
$$P(X_i=1)=\begin{cases} 1\ \  \mathrm{for}\ \  i\leq \lfloor\psi \rfloor -1 \\
\psi+1-\lceil\psi\rceil\ \  \mathrm{for}\ \  i= \lceil\psi \rceil -1\\
0\ \ \mathrm{for}\ \  i\geq \lceil\psi \rceil 
\end{cases}$$
$$P(Y_i=1)=\frac{\psi-1}{M-1}.$$
Now, observe that 
$$P\left(\sum\limits_{i=1}^{M-1}X_i=k-1\right)=[1-|k-\psi|]_{+}$$
and
$$P\left(\sum\limits_{i=1}^{M-1}Y_i=k-1\right)=\binom{M-1}{k-1}\left(\frac{\psi-1}{M-1}\right)^{k-1}\left(\frac{M-\psi}{M-1}\right)^{M-k}$$
for $k=1,...,M$. Since 
$$U-1=\sum\limits_{i=1}^{M-1}Z_i=\sum\limits_{i=1}^{M-1}(DX_i+(1-D)Y_i)=D\sum\limits_{i=1}^{M-1}X_i+(1-D)\sum\limits_{i=1}^{M-1}Y_i$$ 
then
$$P(U=k)=\frac{\rho-C(\psi)}{1-C(\psi)}[1-|k-\psi|]_{+} + \frac{1-\rho}{1-C(\psi)} 
		 \binom{M-1}{k-1}\left(\frac{\psi-1}{M-1}\right)^{k-1}\left(\frac{M-\psi}{M-1}\right)^{M-k}$$
for $k=1,...,M$. The case $\rho<C(\psi)$ is an easy consequence of the fact that $U-1$ has the beta-binomial distribution $BB(M-1,\alpha,\beta)$ with parameters 
 $$\alpha=\frac{(\psi-1)\rho}{(M-1)(C(\psi)-\rho)}, \ \ \ \beta=\frac{(M-\psi)\rho}{(M-1)(C(\psi)-\rho)}.$$		 


\section{Formula for the Gradient of GSD's Log-Likelihood Function}
\label{sec:ap:est}

Denote by $(n_1,..., n_M)$ numbers of observed responses and
$$V'_{\mathrm{min}}(\psi)=-2\psi + \lceil\psi\rceil+\lfloor\psi\rfloor,$$
$$V'_{\mathrm{max}}(\psi)=-2\psi+M+1,$$
$$C'(\psi):=\frac{M-2}{M-1}\ \frac{V_{\mathrm{max}}(\psi)V'_{\mathrm{min}}(\psi)-V'_{\mathrm{max}}(\psi)V_{\mathrm{min}}(\psi)}{(V_{\mathrm{max}}(\psi)-V_{\mathrm{min}}(\psi))^2}.$$
The Log-Likelihood function for $\rho<C(\psi)$ is equal to
\begin{equation}
\begin{split}
l(\psi,\rho)=\sum\limits_{k=1}^M n_k\left[ \log\left(\binom{M-1}{k-1}\right) +
	\sum\limits_{i=0}^{k-2}\log\left(\frac{(\psi-1)\rho}{M-1}+i(C(\psi)-\rho)\right)+\right.\\
	\left. \sum\limits_{i=0}^{M-1-k}\log\left(\frac{(M-\psi)\rho}{M-1}+i(C(\psi)-\rho)\right)-\sum\limits_{i=0}^{M-2}\log\left(\rho+i(C(\psi)-\rho)\right)\right],
\end{split}\nonumber
\end{equation}
and for $\rho\geq C(\psi)$ is equal to
\begin{equation}
\begin{split}
l(\psi,\rho)&=\sum\limits_{k=1}^M n_k \left[\log\Bigg((\rho-C(\psi)) [1-|k-\psi|]_{+} + \right.\\
&\left.(1-\rho)\binom{M-1}{k-1}\left(\frac{\psi-1}{M-1}\right)^{k-1}\left(\frac{M-\psi}{M-1}\right)^{M-k}\Bigg)-\log(1-C(\psi))\right].
\end{split}\nonumber
\end{equation}
The gradient for $\rho<C(\psi)$ is equal to
\begin{equation}
\begin{split}
\frac{\partial l}{\partial \psi}(\psi,\rho)=\sum\limits_{k=1}^M n_k\left[
	\sum\limits_{i=0}^{k-2}\frac{\frac{\rho}{M-1}+i C'(\psi)}{\frac{(\psi-1)\rho}{M-1}+i(C(\psi)-\rho)}+\right.
	\left. \sum\limits_{i=0}^{M-1-k}\frac{-\frac{\rho}{M-1}+i C'(\psi)}{\frac{(M-\psi)\rho}{M-1}+i(C(\psi)-\rho)}-\sum\limits_{i=0}^{M-2}\frac{i C'(\psi)}{\rho+i(C(\psi)-\rho)}\right],
\end{split}\nonumber
\end{equation}

\begin{equation}
\begin{split}
\frac{\partial l}{\partial \rho}(\psi,\rho)=\sum\limits_{k=1}^M n_k\left[
	\sum\limits_{i=0}^{k-2}\frac{\frac{\psi -1}{M-1}-i }{\frac{(\psi-1)\rho}{M-1}+i(C(\psi)-\rho)}+\right.
	\left. \sum\limits_{i=0}^{M-1-k}\frac{\frac{M-\psi}{M-1}-i }{\frac{(M-\psi)\rho}{M-1}+i(C(\psi)-\rho)}+\sum\limits_{i=0}^{M-2}\frac{i-1}{\rho+i(C(\psi)-\rho)}\right],
\end{split}\nonumber
\end{equation}
and for $\rho\geq C(\psi)$ the gradient is equal to
\begin{equation}
\begin{split}
\frac{\partial l}{\partial \psi}(\psi,\rho)&=\sum\limits_{k=1}^M n_k \Bigg[\frac{(\rho-C(\psi))(\mathbb{1}_{[k-1,k]}(\psi)-\mathbb{1}_{[k,k+1]}(\psi))-C'(\psi)[1-|k-\psi|]_{+}}{(\rho-C(\psi)) [1-|k-\psi|]_{+} + 
(1-\rho)\binom{M-1}{k-1}\left(\frac{\psi-1}{M-1}\right)^{k-1}\left(\frac{M-\psi}{M-1}\right)^{M-k}}+ \\
&\frac{\frac{(k-1)(1-\rho)}{M-1}\binom{M-1}{k-1}\left(\frac{\psi-1}{M-1}\right)^{k-2}\left(\frac{M-\psi}{M-1}\right)^{M-k}-\frac{(M-k)(1-\rho)}{M-1}\binom{M-1}{k-1}\left(\frac{\psi-1}{M-1}\right)^{k-1}\left(\frac{M-\psi}{M-1}\right)^{M-1-k}}{(\rho-C(\psi)) [1-|k-\psi|]_{+} + 
(1-\rho)\binom{M-1}{k-1}\left(\frac{\psi-1}{M-1}\right)^{k-1}\left(\frac{M-\psi}{M-1}\right)^{M-k}}  
+ \frac{C'(\psi)}{1-C(\psi)}\Bigg],
\end{split}\nonumber
\end{equation}
\begin{equation}
\begin{split}
\frac{\partial l}{\partial \rho}(\psi,\rho)=\sum\limits_{k=1}^M n_k \frac{[1-|k-\psi|]_{+}-\binom{M-1}{k-1}\left(\frac{\psi-1}{M-1}\right)^{k-1}\left(\frac{M-\psi}{M-1}\right)^{M-k}}{(\rho-C(\psi)) [1-|k-\psi|]_{+} + 
(1-\rho)\binom{M-1}{k-1}\left(\frac{\psi-1}{M-1}\right)^{k-1}\left(\frac{M-\psi}{M-1}\right)^{M-k}}.
\end{split}\nonumber
\end{equation} 


\section{G-Test, Bootstrap Procedure}\label{gtest}
Denote by $(n_1,..., n_M)$ numbers of observed responses, i.e., $n_k$
is the number of responses assigned to response category $k$ and $\sum\limits_{k=1}^{M} n_k=n$.
By $(p_1,..., p_M)$ denote unknown probabilities of the response categories $1,\ldots,M$.
We want to test 
$$H_0:\ (p_1,...,p_M)\ \textrm{are from the GSD}$$ 
against 
$$H_1:\ (p_1,..., p_M)\ \textrm{are not from the GSD.}$$
One should not use the chi-squared test in case of small numbers in selected cells,
i.e., small $n_k$ for some $k\in\{1,\ldots,M\}$. We use a bootstrap version of
the standard likelihood ratio test, i.e., the G-Test. The procedure is as follows:
\begin{enumerate} 
\item Estimate probabilities of the response categories
$(\hat{p}_1,..., \hat{p}_M)$ using the maximum likelihood GSD estimator.
\item Calculate test statistic $T= \sum\limits_{k=1}^{M} n_k \log(n_k / (n \hat{p}_k))$,
where $0 \log(0 / (n \hat{p}_k))=0$.
\item Generate $MC$ (for example, $MC=10\,000$) bootstrap samples of size $n$
from the distribution $(\hat{p}_1,..., \hat{p}_M)$.
Obtain $(m^r_1,..., m^r_M)$, $r=1,\ldots,MC$, where $m^r_k$ is the
number of responses assigned to response category $k$ in the $r$-th bootstrap sample.
\item Estimate probabilities of the response categories
$(\hat{q}^r_1,..., \hat{q}^r_M)$ for every
bootstrap sample $(m^r_1,..., m^r_M)$ using the maximum likelihood GSD estimator. 
\item Calculate bootstrap statistics
\begin{equation*}
T_r= \sum\limits_{k=1}^{M} m^r_k \log(m^r_k / (n \hat{q}^r_k)),
\end{equation*}
where $0 \log(0 / (n \hat{q}^r_k))=0$. 
\item Calculate bootstrap $p$-value using the following equation
\begin{equation*}
p=\frac{1}{MC}\sum\limits_{r=1}^{MC} I(T_r\geq T),
\end{equation*}
where $I(x)$ is one if $x$ is true or 0 if $x$ is false.
\end{enumerate}


\section{Modified EPMF and GSD}
\label{app:parameters_estimation_modification}
To resolve the problem of empty cells for the empirical distribution, we can simply add $0.5$ to all response category counts (cf. \cite{pagano2018principles}), i.e.,
$$
\forall k \in \{ 1, \cdots, M \}, \quad \hat{v}_k=\frac{n_k+0.5}{n+\frac{M}{2}},
$$
where $n_k$ is the response count of category $k$ in a bootstrap sample.

For the GSD it is enough to estimate parameters $\psi,\rho$ on the set $[1+\epsilon_{\psi_d}(n),5-\epsilon_{\psi_u}(n)]\times[0+\epsilon_{\rho_{\mathrm{d}}}(n),1-\epsilon_{\rho_{\mathrm{u}}}(n)]$, where $\epsilon_{\psi_{\cdot}}(n)>0$, $\epsilon_{\rho_{\cdot}}(n)>0$ and $\lim\limits_{n \rightarrow \infty} \epsilon_{\psi_{\cdot}}(n)=\lim\limits_{n \rightarrow \infty} \epsilon_{\rho_{\cdot}}(n)=0$.

To define $\epsilon_{\psi_{\cdot}}(n)$ and $\epsilon_{\rho_{\cdot}}(n)$ we introduce
a limit for the maximum probability any two response categories can add up to (and call
it $p_{\max}$). Importantly, when assessing $p_{\max}$, we only take into account two
most probable response categories. This can be formally written as follows:
\begin{equation}\label{eq:pmax}
 p_{\max} = \max_{(i,j)\in\{1,...,M\}^2: i\neq j} P(U=i) + P(U=j) 
\end{equation}

The final algorithm for fitting the GSD to a sample is as follows. Find such
($\hat{\psi}$, $\hat{\rho}$) that satisfies the following two criteria:
\begin{enumerate}
    \item $p_{\max} \leq 1 - \frac{1}{n}$, where $p_{\max}$ is given by equation (\ref{eq:pmax}) and $n$ is the sample size, and
    \item the likelihood function has the maximum value.
\end{enumerate}

\begin{figure}
	\centering
	\includegraphics[width=0.7\textwidth]{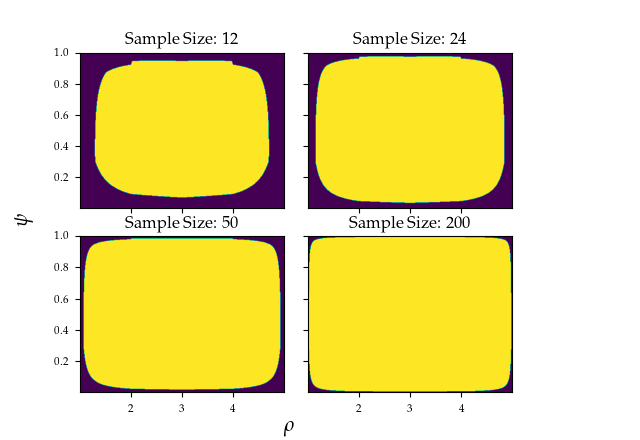}
	\caption{Boundary for $\psi$ and $\rho$ for a given sample size $n$ and $p_{\max}\leq 1 - \frac{1}{n}$. Yellow color marks ($\psi$, $\rho$) pairs considered in the MLE algorithm.}
	\label{fig:psi_rho_boundary}
\end{figure}

An example of $\psi$ and $\rho$ ranges for different sample sizes and $M=5$ is shown in Fig.~\ref{fig:psi_rho_boundary}. 

\section{Mulitidimensional Case}
\label{app:multi}
Let us consider a multidimensional model of responses with $n$ raters (also referred to as subjects) and $m$ objects (also referred to as stimuli), i.e.,
$$U_{ij}=\psi_j + \epsilon_{ij}, \ \ \ i\in{1,...,n},\ j\in{1,...,m}$$
where $\epsilon_{ij}+\psi_j$ has the GSD$(\psi_j,\rho_i)$ distribution. In this model, every object (e.g., a video) has its own quality $\psi_j$ and every rater has their own confidence parameter $\rho_i$. 

For numerical experiments, we generated $100\,000$ response matrices $U_{i,j}$ according to the following generative process: $\psi_j\sim \text{Uniform}(1,5)$, $\rho_i\sim \text{Uniform}(0,1)$.
For each sample a probabilistic matrix factorisation $U_{ij}~\sim GSD(\hat\psi_j,\hat\rho_i)$ done by MLE yields recovered estimates $\hat\psi_j$ and $\hat\rho_i$.

\begin{figure}
	\begin{center}
        \includegraphics[scale=0.7]{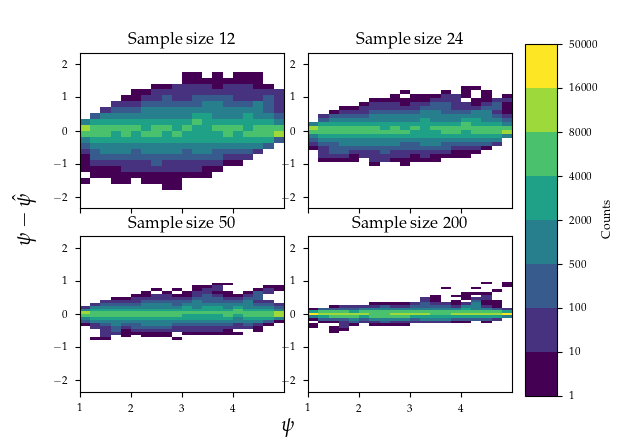}
		\caption{Estimation accuracy for $\psi$ estimation for the multidimensional model with $n$ subjects scoring $m=n$ objects. 
		} \label{fig:RMSD_psi_multi_dim}
	\end{center}
\end{figure}

\begin{figure}
	\begin{center}
        \includegraphics[scale=0.7]{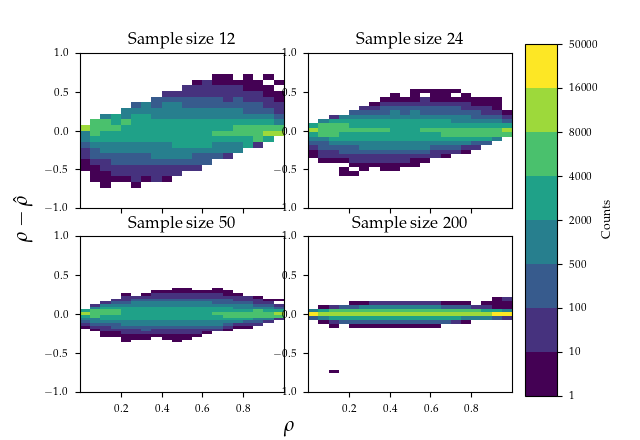}
		\caption{Estimation accuracy for $\rho$ estimation for the multidimensional model with $n$ subjects scoring $m=n$ objects. 
		} \label{fig:RMSD_rho_multi_dim}
	\end{center}
\end{figure}

In Fig.~\ref{fig:RMSD_psi_multi_dim} we present the estimation accuracy for a fixed $\psi$ of one object in a multidimensional numerical experiment for $n=m=12, 24, 50, 200$. All other parameters where random (uniformly distributed). For estimation, we used the gradient based method, using the formulae for the gradient of GSD log-likelihood function from Appendix \ref{sec:ap:est}. In Fig.~\ref{fig:RMSD_rho_multi_dim} we present estimation accuracy for fixed $\rho$ of one rater in the same multidimensional numerical experiment. As one can see, our multidimensional estimator of GSD parameters is very accurate even for relatively small sample sizes. Notice that in case of $n=m=200$, we estimate $400$ parameters using the MLE, i.e., we estimate $\psi_1,...,\psi_{200},\rho_1,...,\rho_{200}$. The accuracy of the estimator is higher if both $n$ and $m$ are larger.       

\end{document}